# Activity-dependent glassy cell mechanics II：

## Non-thermal fluctuations under metabolic activity

K. Umeda, K. Nishizawa, W. Nagao, S. Inokuchi, Y. Sugino, H. Ebata, and *D. Mizuno

1. Department of Physics, Kyushu University, 819-0395 Fukuoka, Japan.
2. Institute of Developmental Biology of Marseille, Campus de Luminy case 907, 13288 Marseille Cedex 09, France.

* Correspondence: mizuno@phys.kyushu-u.ac.jp; Tel.: +81-92-802-4092

**running title**：**mesoscale dynamics of glassy cytoplasm**

**Abstract:** The glassy cytoplasm, crowded with bio-macromolecules, is fluidized in living cells by mechanical energy derived from metabolism. Characterizing the living cytoplasm as a non-equilibrium system is crucial in elucidating the intricate mechanism that relates cell mechanics to metabolic activities. In this study, we conducted active and passive microrheology in eukaryotic cells, and quantified non-thermal fluctuations from the violation of the fluctuation-dissipation theorem (FDT). The power spectral density corresponding to active force generation was then estimated following the Langevin theory extended to non-equilibrium systems. Experiments performed while regulating cellular metabolic activity showed that the non-thermal displacement fluctuation, rather than the active non-thermal force, directly correlates with metabolism. We discuss how mechano-enzymes in living cells do not act as a collection of microscopic objects; rather, they generate meso-scale collective fluctuations that directly correlate with enzymatic activity. The correlation is lost at long time scales because of the mesoscopic structural relaxations induced by the metabolic activities. Since the efficiency with which energy is converted to non-thermal fluctuations decreases as the cytoplasm becomes fluidic, the fluidization of the cytoplasm stops at critical jamming. Regardless of the presence or absence of structural relaxations in the cytoplasm, we demonstrate that non-thermal fluctuations in a probe particle can serve as a valuable indicator of those metabolic activities that typically perturb the mechanical environment within cells.

**Statement of Significance:** A living cell's interior is crowded with various enzymes and organelles that incessantly generate non-thermal fluctuations by utilizing the energy released during ATP hydrolysis. In this study, the mechanics of the living cytoplasm were investigated by observing the non-thermal fluctuation as characterized by the violation of the fluctuation-dissipation theorem. The glassy intracellular cytoplasm is fluidized by these fluctuations such that they approach the critical jamming state. By focusing on the meso-scale nature of non-thermal fluctuations, we explain the enigma of intracellular mechanics, *i.e.*, the power-law rheology $G(\omega) \propto (-i\omega)^{1/2}$ typical of the critical jamming state, the time-scale separation of active and thermal diffusion, *etc*. By presenting a continuum model to describe mesoscale intracellular dynamics, we show that the non-thermal fluctuation correlates with the metabolic activity of cells.

## INTRODUCTION

The mechanical properties of cells play a crucial role in determining their functions and behaviors. The cytoskeleton, a complex network of semi-flexible biopolymers, has been widely recognized as the primary determinant of cell mechanics (1, 2). This is because conventional experimental techniques *e.g.*, AFM, magnetic twisting cytometry, micropipette aspiration have mostly focused on measuring the mechanical properties of the cell surface where the actin cytoskeleton is densely expressed (2-6). Results of these studies thus conformed to theoretical predictions for semi-flexible polymer networks (2, 4). Inside of cells, however, other bio-macromolecules are densely expressed in the interstices of the sparse cytoskeletal network (7, 8). The mechanics of the crowded cytoplasm are equally important, as they govern the dynamics of various physiological processes. It was suggested that the cytoplasm may possess glass-forming capabilities, which in fact impacts cell mechanics (9). A cytoplasm that lacks metabolic activities solidifies in a similar manner to a glass transition (5, 10, 11) because cell interior is highly crowded with various colloidal and polymeric components (12). On the other hand, metabolic activities in living cells seem to prevent vitrification (11-14).

Living cells are inherently out of equilibrium driven by the activity of functional bio-macromolecules such as molecular motors that harness chemical energy from ATP hydrolysis to perform mechanical work (15-17). Not only the molecular motors, but other enzymes and organelles containing active enzymes, collectively referred to as mechano-enzymes in this article, may also perturb the cell environment while catalyzing biochemical reactions (18, 19). Consequently, non-thermal fluctuations are ubiquitous in metabolically active cells while thermal fluctuations are suppressed because the crowded cytoplasm is highly viscoelastic (12, 20). Fluctuations have a profound impact on the mechanical properties of cells, which are made of soft materials. The elasticities of colloidal glasses and semiflexible polymer networks, which are physical models for cell mechanics, are derived from entropy, with thermal fluctuations playing a pivotal role (21, 22). In thermodynamic equilibrium, temperature, fluctuations, and mechanics are tightly related to each other by the fluctuation-dissipation theorem (FDT) (23, 24). While non-thermal fluctuations in cells are not bound by equilibrium physics, they modulate cell mechanics as they do in other soft materials (25, 26). This is attributed to the anomalous statistics and spatiotemporal dynamics of non-thermal fluctuations (27). An important example is “active fluidization”; the glassy cytoplasm becomes fluidized during ordinary metabolism in living cells (11). Therefore, characterizing non-thermal fluctuations is crucial for elucidating the non-equilibrium mechanics of the active cytoplasm.

Microrheology (MR) is a powerful tool for investigating the mechanics and dynamics of a medium at length scales of approximately a micrometer, accomplished by tracking the motion of micron-sized tracer beads imbedded in a specimen (28, 29). Active microrheology (AMR) (30, 31) involves applying an external force to a tracer bead and measuring its response displacement. By applying a sinusoidal force, we obtain the complex shear viscoelastic modulus $G(\omega)$ of the surrounding medium as a function of the angular frequency $\omega$. Passive microrheology (PMR) involves measuring the spontaneous fluctuations of the imbedded beads (30, 32, 33). FDT applies to PMR in specimens under thermodynamic equilibrium (23), which allows us to obtain $G(\omega)$ of the medium. AMR and PMR therefore provide equivalent information. However, living cells are driven far from equilibrium by metabolic activities of mechano-enzymes, and thus the FDT does not apply to total fluctuations in living cells (20, 26, 34-36). However, the strength of thermal fluctuations can still be estimated from AMR by assuming FDT, enabling the evaluation of non-thermal fluctuations by subtracting the estimated power of thermal fluctuations from the total fluctuations measured with PMR. The non-thermal fluctuations manifest how far the sample is driven from equilibrium (16, 25, 30, 37-40). Therefore, MR experiments that simultaneously conduct AMR and PMR have the unique potential to investigate active mechanics, *i.e.* the profound non-equilibrium relation between activeness and mechanics in living cells (26).

In the study of living cells, both AMR and PMR can be conducted simultaneously by applying an optical-trapping force to the probe particle and measuring its displacements using back-focal-plane laser interferometry (BFPI) (30, 41). However, due to the intense non-thermal fluctuations in living cells, the probe particle can easily move out of the laser focus during the experiment. To overcome this limitation, a feedback-tracking MR technique was developed using an optical-trap-based MR implemented with a three-dimensional feedback-controlled sample stage (20, 42). This approach enables stable tracking of a fluctuating probe in living cells. In this study, we will utilize this MR technique to investigate the intracellular mechanics of living HeLa cells while simultaneously controlling their metabolic activities.

We employed feedback-tracking MR to investigate the active mechanics of HeLa cells cultured in an epithelial-like monolayer sheet. Our previous study reported the mechanical properties of ordinary HeLa cells measured using AMR. $G(\omega)$ in ordinary cells exhibited a frequency dependency $G(\omega) \propto (-i\omega)^{1/2}$ in the whole range of frequencies measured (0.1 Hz ~ 10 kHz) (12, 20). On the other hand, a static elasticity $G_0$ emerged in ATP-depleted HeLa cells as $G(\omega) = G_0 + A(-i\omega)^{1/2}$ (13). In this study, we aimed to investigate the physical mechanism of the regulation in terms of non-thermal fluctuations. By utilizing the violation of the fluctuation-dissipation theorem (FDT) (20, 26, 37), we quantified non-thermal fluctuations in the probe particle's displacement. Non-thermal fluctuations were significantly reduced whereas viscoelasticity was increased in ATP-depleted cells. These results were consistent with the concept of a glassy cytoplasm, where the mechanical energy supplied by the non-thermal fluctuations induces structural relaxations and fluidizes the glassy cytoplasm.

To investigate the mechanism of active fluidization, we utilized the Langevin model to estimate force generation and/or mechanical energy converted by mechano-enzymes in non-equilibrium systems (30, 39, 40). Contrary to expectations, we found that the active force generation did not reflect the cell's metabolic activity whereas the non-thermal displacement fluctuations did. Our analysis suggests that mechano-enzymes do not behave like microscopic objects such as solvent molecules. Instead, they produce large-scale collective (hereafter, referred to as "mesoscopic") fluctuations which directly correlate with the activity of mechano-enzymes and impact probe particle dynamics. When the medium is fluidized, energy transfer from mechano-enzymes to non-thermal fluctuations is inefficient. This implies that the cytoplasm of living cells only approaches critical jamming, indicated by the characteristic power-law rheology $G(\omega) \propto (-i\omega)^{1/2}$, but does not reach a more fluid state.

The highly viscoelastic mechanical properties indicate that the cytoplasm is not completely fluid. Thermal fluctuations are expected to be anomalous (sub-diffusive) by assuming the FDT. Fluid dynamics, *i.e.*, diffusion, would be expected at lower frequencies than those we measured. However, non-thermal fluctuations exhibited diffusion-like dynamics within the measured frequency range. Mesoscopic structural relaxations induced by metabolic activities seem to erase the correlation between the active force and mesoscopic non-thermal fluctuations. Nonetheless, we qualitatively discuss how the strength of non-thermal fluctuations reflects overall active metabolism of living cells, regardless of the presence or absence of structural relaxations.

## MATERIALS AND METHODS

### Cell preparation

HeLa cells were seeded on fibronectin-coated glass-bottom petri dishes in Dulbecco's modified Eagle's medium (Wako, D-Mem, high glucose) with glucose (1 mg/ml), penicillin (100 U/ml), streptomycin (0.1 mg/ml), amphotericin B (250 mg/ml), and 10% fetal bovine serum (FBS) at 37°C. Cells were cultured in a $CO_2$ incubator until they formed a confluent epithelial-like monolayer sheet. The probe particles (melamine particles 1 μm diameter, micro Particles GmbH) coated with polyethylene glycol (PEG) strands (43) were introduced into the HeLa cells using a gene gun (PDS-1000/He, Bio-Rad). PEG coating has been widely used to generally passivate probe surfaces to biomaterials. In aqueous environments, hydrophilic PEG acts as a polymer brush and prevents sticking to other objects or to other molecules. Excess beads that did not enter the cells were removed by washing the dishes with phosphate-buffered saline. After replacing the medium with fresh medium, the dishes were placed in an $CO_2$ incubator at least overnight until any cell damage was repaired. Supplying $CO_2$ in the MR setup causes additional noise to appear in the experimental data. Therefore, MR experiments were performed in a $CO_2$-independent culture medium (Gibco, L-15 medium) with 10% FBS serum. All measurements were performed at 37°C. FBS was not included in the media for MR experiments under ATP depletion.

The high energy molecule ATP is mainly produced in cells via two metabolic pathways, glycolytic in the cytoplasm and oxidative phosphorylation in the mitochondria. In order to deplete intracellular ATP, cells were cultured in a nutrient free medium (D-Mem without glucose and FBS) with 50 mM 2-deoxy-D-glucose and 10 mM sodium azide (4). 2-deoxy-D-glucose and sodium azide inhibit glycolysis and oxidative ATP production, respectively. Cells were incubated at 37 °C for ~10 hours prior to MR measurement so that cells consume the ATP stored in cells. The medium was then replaced with a $CO_2$-independent culture medium (L-15 medium with 50 mM 2-deoxy-D-glucose, 10 mM sodium azide, without FBS) to perform microrheology experiments. Intracellular ATP was measured using a CellTiter-Glo® Luminescent Cell Viability Assay kit (Promega, G7570) according to the manufacturer's instructions (13).

**MR theory in living cells**

Details of the MR experiments in living cells under feedback control are given elsewhere (20, 42). Below, we will summarize the theoretical foundations of MR which are necessary to discuss our MR data.

*Conventional MR at equilibrium*

In the case of AMR, a small sinusoidal force $F(t)=\hat{F}(\omega)e^{-i\omega t}$ was applied to the probe particle, by optically trapping it with the drive laser ($\lambda$ = 1064 nm, Fig. 1A). The focus position of the drive laser was oscillated with an acousto-optic deflector. The probe displacement synchronous to the applied force $u(t)=\hat{u}(\omega)e^{-i\omega t}$ was measured by back-focal-plane laser interferometry (BFPI) (41) using a quadrant photodiode (QPD) and a probe laser ($\lambda$ = 830 nm). Here, "ˆ" denotes the magnitude of the sinusoidal signal, expressed in polar form. The frequency response function $\tilde{\alpha}(\omega)=\alpha'(\omega)+i\alpha''(\omega)\equiv\langle\hat{u}(\omega)\rangle/\hat{F}(\omega)$ is then obtained as a complex quantity since $u(t)$ is out of phase with $F(t)$ because of the time delay. The shear viscoelastic modulus $G(\omega)$ of the surrounding media is also obtained as a complex quantity via the generalized Stokes relation applied to the sinusoidal response,

$$G(\omega)=G'(\omega)-iG''(\omega)=\frac{1}{6\pi a\tilde{\alpha}(\omega)}. \qquad (1)$$

where $a$ is a radius of a spherical probe particle.

For PMR, the displacement $u(t)$ of a probe particle was measured without external perturbations (Fig. 1B). The thermal fluctuation $u_{\mathrm{th}}(t)$ is given by the over-damped Langevin equation,

$$\int_{-\infty}^{t}\gamma(t-t')\dot{u}_{\mathrm{th}}(t')dt'=f_{\mathrm{th}}(t)\Leftrightarrow u_{\mathrm{th}}(t)=\int_{-\infty}^{t}\alpha(t-t')f_{\mathrm{th}}(t')dt' \qquad (2)$$

where $f_{\mathrm{th}}(t)$ is the thermally fluctuating force, $\gamma(t)$ is the friction function and $\alpha(t)$ is the impulse response function. Here we provide two equivalent expressions for later convenience. The Fourier transforms of $\gamma(t)$ and $\alpha(t)$ are related to each other by

$$\tilde{\gamma}(\omega)=-1/[i\omega\tilde{\alpha}(\omega)]. \qquad (3)$$

"˜" denotes the Fourier transform of a function in this study. The fluctuation-dissipation theorem (FDT) of the 1st kind relates $u_{\mathrm{th}}(t)$ to $\tilde{\alpha}(\omega)$ as (23, 32)

$$\left\langle |\tilde{u}_{\mathrm{th}}(\omega)|^2 \right\rangle = \frac{2k_{\mathrm{B}}T}{\omega}\alpha''(\omega). \tag{4}$$

Here, $\langle\ \rangle$ indicates the statistical average and the left-hand side of Eq. (4) indicates the power spectral density (PSD) of thermal probe displacements $u_{\mathrm{th}}(t)$. $k_{\mathrm{B}}$ and $T$ are the Boltzmann constant, and absolute temperature, respectively. Eq. (2) and (4) are consistent because of the FDT of the 2nd kind,

$$\left\langle |\tilde{f}_{\mathrm{th}}(\omega)|^2 \right\rangle = 2k_{\mathrm{B}}T\gamma'(\omega), \tag{5}$$

where $\tilde{\gamma}(\omega) = \gamma'(\omega) + i\gamma''(\omega)$.

*FDT violation far from equilibrium*

In living cells, mechano-enzymes convert chemical energy into a mechanical form, producing non-thermal active fluctuations in the living cytoplasm. The spontaneous motion of the probe particle $u(t)$ measured with PMR is comprised of the active non-thermal fluctuation $u_{\mathrm{A}}(t)$ and the thermal probe fluctuation $u_{\mathrm{th}}(t)$ as $u(t) = u_{\mathrm{th}}(t) + u_{\mathrm{A}}(t)$. Under the assumption that $u_{\mathrm{th}}(t)$ and $u_{\mathrm{A}}(t)$ are not correlated, $\left\langle |\tilde{u}(\omega)|^2 \right\rangle = \left\langle |\tilde{u}_{\mathrm{A}}(\omega)|^2 \right\rangle + \left\langle |\tilde{u}_{\mathrm{th}}(\omega)|^2 \right\rangle$ is greater than $\left\langle |\tilde{u}_{\mathrm{th}}(\omega)|^2 \right\rangle = 2k_{\mathrm{B}}T\alpha''(\omega)/\omega$ as schematically shown in Fig. 1C. This is commonly referred to as violation of the FDT. The degree of FDT violation is equivalent to the PSD of the active fluctuations,

$$\left\langle |\tilde{u}_{\mathrm{A}}(\omega)|^2 \right\rangle = \left\langle |\tilde{u}(\omega)|^2 \right\rangle - \frac{2k_{\mathrm{B}}T}{\omega}\alpha''(\omega). \tag{6}$$

The FDT violation can be also presented with the PSD of velocities, $v(t) = \dot{u}(t), v_{\mathrm{th}} = \dot{u}_{\mathrm{th}}(t), v_{\mathrm{A}} = \dot{u}_{\mathrm{A}}(t)$ as shown in Fig. 1D. As guaranteed by Parseval's theorem, the total power $Q$ of non-thermal fluctuations is then obtained by an integration over frequency,

$$Q \equiv \frac{1}{2\pi}\int_{-\infty}^{\infty} \left\langle |\tilde{u}_{\mathrm{A}}(\omega)|^2 \right\rangle d\omega = \frac{1}{\pi}\int_{0}^{\infty} \left\langle |\tilde{u}_{\mathrm{A}}(\omega)|^2 \right\rangle d\omega \tag{7}$$

When estimating $Q$ from actual MR data, the integration on the right hand side of Eq. (7) can only be conducted over a limited range of frequencies. In practice, $Q$ is calculated from $N$ discrete data sampled at an interval $\Delta t = t_{\mathrm{e}}/N$ as,

$$Q(t_{\mathrm{e}}) \equiv \frac{1}{\pi}\sum_{n=1}^{N/2} \left\langle |\tilde{u}_{\mathrm{A}}(n\Delta\omega)|^2 \right\rangle \Delta\omega \sim \frac{1}{\pi}\int_{\omega_{\mathrm{L}}}^{\omega_{\mathrm{H}}} \left\langle |\tilde{u}_{\mathrm{A}}(\omega)|^2 \right\rangle d\omega\ , \tag{8}$$

where $\Delta\omega = 2\pi/t_{\mathrm{e}}$ . $\omega_{\mathrm{H}} = \pi/\Delta t$ and $\omega_{\mathrm{L}} \sim 1/t_{\mathrm{e}}$ are the high- and low-frequency cutoffs. Given that $\left\langle |\tilde{u}_{\mathrm{A}}(\omega)|^2 \right\rangle$ decays with $\omega$, *e.g.*, $\left\langle |\tilde{u}_{\mathrm{A}}(\omega)|^2 \right\rangle \propto \omega^{-2}$ for the intracellular fluctuation reported in this study, $Q(t_{\mathrm{e}})$ is unaffected by the high-frequency cutoff, but is dependent on the low frequency cutoff which is determined by the duration of the experiment $t_{\mathrm{e}}$. Hence, all experiments were performed with the same $t_{\mathrm{e}}$, to compare between cells with varying metabolic activities.

*Langevin model extended to non-equilibrium situation*

FDT violation has been investigated by extending the Langevin model expressed in Eq. (2) to non-equilibrium situations (30, 35, 39, 40). It has been hypothesized that the active fluctuation $u_{\mathrm{A}}(t)$ is driven by an additional active force $f_{\mathrm{A}}(t)$ acting on the probe particle. Assuming that $u_{\mathrm{A}}(t)$ linearly responds to $f_{\mathrm{A}}(t)$ via the same kernels [$\gamma(t)$ and $\alpha(t)$] such that,

$$\int_{-\infty}^{t} \gamma(t-t')\dot{u}_{\mathrm{A}}(t')dt' = f_{\mathrm{A}}(t) \Leftrightarrow u_{\mathrm{A}}(t) = \int_{-\infty}^{t} \alpha(t-t') f_{\mathrm{A}}(t')dt', \qquad (9)$$

the Langevin equation [Eq. (2)] can be extended to non-equilibrium situations, *i.e.*,

$$\int_{-\infty}^{t} \gamma(t-t')\{\dot{u}_{\mathrm{A}}(t') + \dot{u}_{\mathrm{th}}(t')\}\, dt' = f_{\mathrm{A}}(t) + f_{\mathrm{th}}(t) \cdot \qquad (10)$$

Here, "$\dot{}$" indicates the time derivative. The PSD of the active force $f_{\mathrm{A}}(t)$ is then obtained via the formula (25, 30, 39)

$$\left\langle |\tilde{f}_{\mathrm{A}}(\omega)|^2 \right\rangle = \omega^2 |\tilde{\gamma}(\omega)|^2 \cdot \left\langle |\tilde{u}_{\mathrm{A}}(\omega)|^2 \right\rangle = |6\pi a G(\omega)|^2 \cdot \left\langle |\tilde{u}_{\mathrm{A}}(\omega)|^2 \right\rangle. \qquad (11)$$

Under the same assumption, the mechanical power of the mechano-enzyme (the work done by $f_{\mathrm{A}}(t)$ per unit time) is obtained by the Harada-Sasa equation as (16, 38, 44)

$$\langle v_{\mathrm{A}}(t) \cdot f_{\mathrm{A}}(t) \rangle = \frac{1}{2\pi} \int_{-\infty}^{\infty} \gamma'(\omega) \left\langle |\tilde{v}_{\mathrm{A}}(\omega)|^2 \right\rangle d\omega. \qquad (12)$$

Here, $\tilde{v}_{\mathrm{A}}(\omega) = -i\omega \tilde{u}_{\mathrm{A}}(\omega)$ is the Fourier transform of the active probe velocity $v_{\mathrm{A}}(t)$ whose PSD is assessed by

$$\left\langle |\tilde{v}_{\mathrm{A}}(\omega)|^2 \right\rangle = \left\langle |\tilde{v}(\omega)|^2 \right\rangle - 2\omega k_{\mathrm{B}} T \alpha''(\omega). \qquad (13)$$

where the FDT $\left\langle |\tilde{v}_{\mathrm{th}}(\omega)|^2 \right\rangle = 2\omega k_{\mathrm{B}} T \alpha''(\omega)$ is used to estimate the thermal velocity fluctuation. The right-hand side of Eq. (11) and (12) can be calculated by using the data obtained from AMR and PMR experiments.

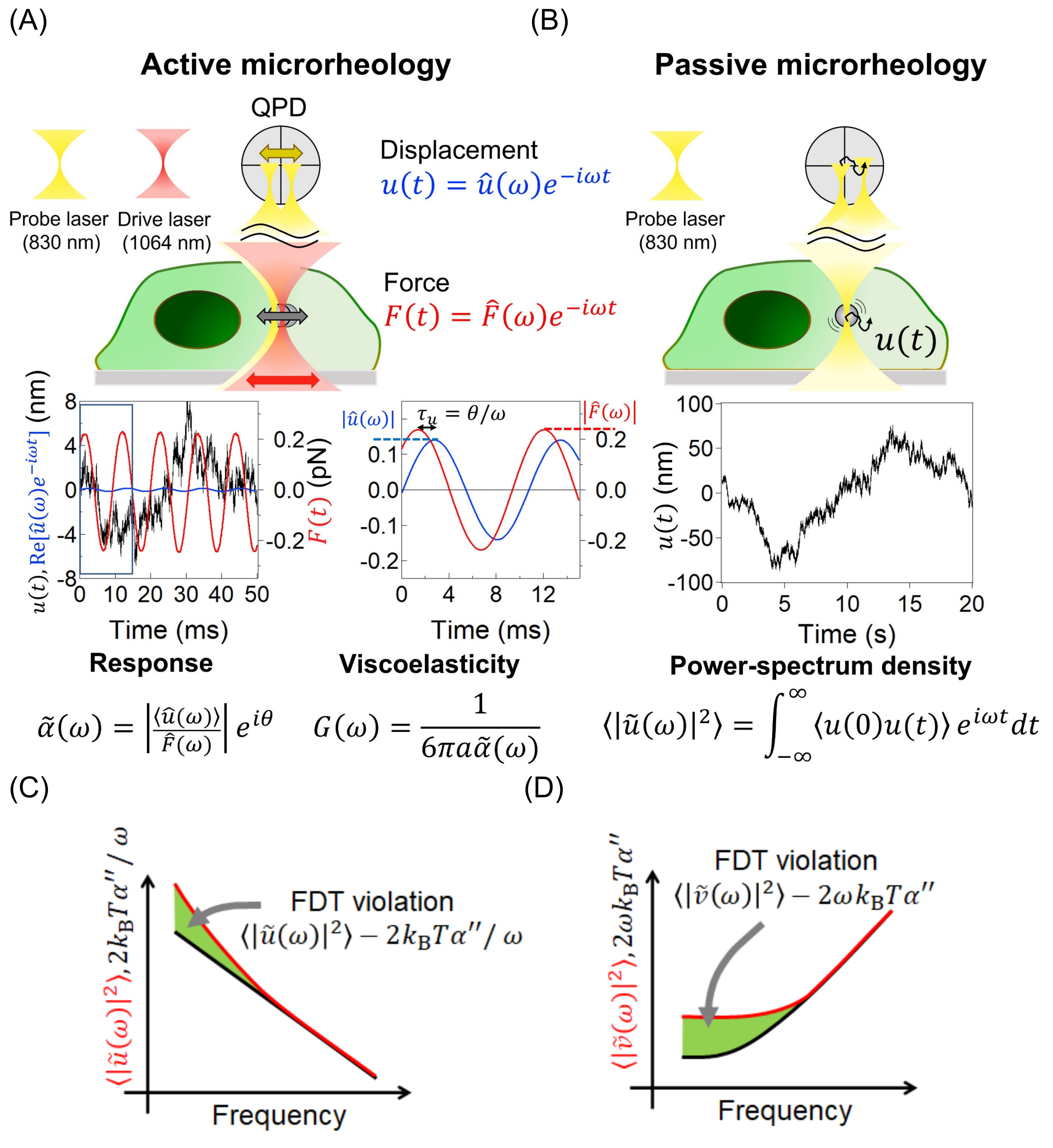

FIGURE 1 Schematics of AMR, PMR and FDT violation. (A) In AMR, a sinusoidal external force is applied to a probe particle using a drive laser (λ = 1064 nm). The probe particle's displacement is measured using a probe laser (λ = 830 nm). Whereas the minute response to the applied force is hidden behind (non-) thermal fluctuations, a lock-in amplifier detects the phase delay $\theta = \tau_u\omega$ and amplitude of the response $|\langle\hat{u}(\omega)\rangle|$, yielding the response function $\alpha(\omega)$ and viscoelasticity $G(\omega)$. (B) In passive MR, spontaneous probe fluctuations are measured using an 830 nm laser. By calculating the power spectral density of the fluctuation, the fluctuation-dissipation theorem [$\langle|\tilde{u}(\omega)|^2\rangle = 2k_BT\alpha''(\omega)/\omega$ in (C) and $\langle|\tilde{v}(\omega)|^2\rangle = 2\omega k_BT\alpha''(\omega)$ in (D)] is directly tested. In out-of-equilibrium systems where FDT is violated, the power-spectrum density of active fluctuations is obtained as $\langle|\tilde{u}_A(\omega)|^2\rangle = \langle|\tilde{u}(\omega)|^2\rangle - 2k_BT\alpha''(\omega)/\omega$ and $\langle|\tilde{v}_A(\omega)|^2\rangle = \langle|\tilde{v}(\omega)|^2\rangle - 2\omega k_BT\alpha''(\omega)$, respectively.

# RESULTS

## FDT violation and ATP-dependent glassy mechanical properties of cytoplasm

Using the feedback-tracking technique, AMR and PMR were performed in HeLa cells derived from human cervical cancer. In order to investigate the dependency on metabolic activities, experiments were conducted with and without ATP depletion (4). Melamine particles ($2a$ = 1 μm) were incorporated into cells forming an epithelial-like confluent monolayer on the surface of a glass-bottom dish. Particles incorporated at the center of cells between the cell membrane and the nuclear membrane were used as probes. The PSD of the thermal fluctuation $\langle |\tilde{u}_{\mathrm{th}}(\omega)|^2 \rangle = 2k_{\mathrm{B}}T\alpha''(\omega)/\omega$ (black circles, AMR) and the total fluctuation $\langle |\tilde{u}(\omega)|^2 \rangle$ (red circles, PMR) are shown in Fig. 2. In order to investigate the metabolism dependency, cells were incubated in a medium which causes ATP depletion. The averaged values with (n = 9) and without ATP depletion (n = 21) are shown in Fig. 2A and B, respectively. In this study, averages and SDs were calculated for the logarithm of experimental results since the logarithms of MR data in cells are approximately distributed with Gaussian (4). $\langle |\tilde{u}(\omega)|^2 \rangle$ agrees with $\langle |\tilde{u}_{\mathrm{th}}(\omega)|^2 \rangle = 2k_{\mathrm{B}}T\alpha''(\omega)/\omega$ at frequencies higher than 10 Hz, indicating that the FDT [Eq. (4)] is satisfied. On the other hand, $\langle |\tilde{u}(\omega)|^2 \rangle$ is remarkably greater than thermal effects as estimated from AMR at lower frequencies, indicating FDT violation $\langle |\tilde{u}(\omega)|^2 \rangle - 2k_{\mathrm{B}}T\alpha''(\omega)/\omega > 0$. The green-colored region indicates the non-thermal fluctuation $\langle |\tilde{u}_{\mathrm{A}}(\omega)|^2 \rangle$ given in Eq. (6).

Before investigating FDT violation, we go over the metabolism dependency of the dynamic shear viscoelastic modulus $G(\omega)$ measured with AMR (13). $G(\omega)$ measured with AMR is shown before and after ATP depletion in Fig. 2C and D, respectively. The frequency dependency $G(\omega) \propto (-i\omega)^{1/2}$ was observed over the entire frequency range measured (0.1 Hz ~ 10 kHz). As reported in a prior study (13, 20), epithelial cells are roughly as tall as they are wide when confined in confluent monolayers. Actin cytoskeletons are less expressed inside such epithelial cells (13, 45-48). Also, the probe particles were coated with polyethylene glycol polymers, inhibiting adherence to objects in cells, including the cytoskeleton. Under such experimental conditions, crowding by various bio-macromolecules plays a greater role in determining intracellular mechanics. ATP-depleted cells became elastic; the real part $G'(\omega)$ increased at low frequencies and exhibited an elastic plateau $G_0$ which is constant over frequency. Note that apart from the emergence of the static elasticity, the power-law component $G(\omega) \propto (-i\omega)^{1/2}$ was also present in ATP-depleted cells such that $G(\omega) = G_0 + A(-i\omega)^{1/2}$. Therefore, our results indicate that a cytoplasm with less cytoskeleton stiffens with ATP-depletion in a metabolism-dependent manner. Though the results in this study are limited for epithelial cells in confluent monolayer, we are getting similar results for different cells and conditions, which will be reported elsewhere.

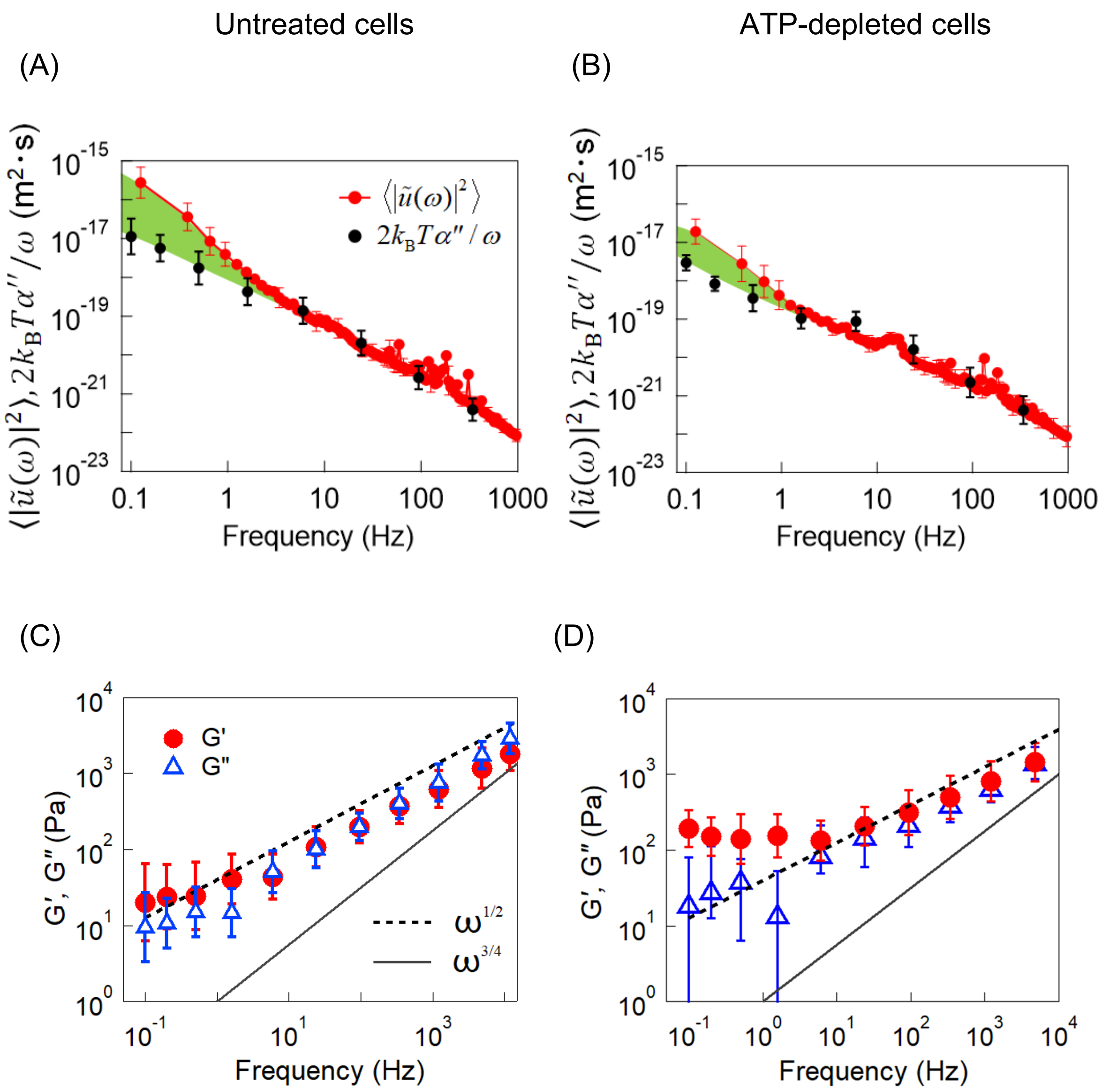

FIGURE. 2
FDT violation in (A) untreated (n = 21) and (B) ATP-depleted HeLa cells (n = 9). The difference between $\langle|\tilde{u}(\omega)|^2\rangle$ measured with PMR (red circles) and $\langle|\tilde{u}_{\mathrm{th}}(\omega)|^2\rangle = 2k_{\mathrm{B}}T\alpha''(\omega)/\omega$ measured with AMR (black circles), shown by the green colored region at low frequencies, indicates non-thermal fluctuation $\langle|\tilde{u}_{\mathrm{A}}(\omega)|^2\rangle$. The peaks observed at approximately 100 Hz and 20 Hz in the PMR spectrum are likely due to mechanical or electronic noise. Complex shear modulus $G(\omega)$ of (C) untreated and (D) ATP-depleted cells. Red circles and blue triangles indicate the real and imaginary part, respectively. Broken and solid lines indicate the frequency dependency typical of glassy cytoplasms ($\propto \omega^{1/2}$) and cytoskeletons ($\propto \omega^{3/4}$), respectively.

**Non-thermal fluctuation correlates with metabolic activities**

To elucidate the mechanism behind FDT violation, we now investigate the probe fluctuation in cells in more detail. Firstly, the power spectral density of the active forces $\langle|\tilde{f}_{\mathrm{A}}(\omega)|^2\rangle$ were obtained using Eq. (11). $\langle|\tilde{f}_{\mathrm{A}}(\omega)|^2\rangle$ calculated for each probe particle and their statistical average are shown in Fig. 3A and B for untreated and ATP-depleted cells, respectively. Results were fitted with the power-law function $\langle|\tilde{f}_{\mathrm{A}}(\omega)|^2\rangle = B\omega^{\beta}$ with $\beta = -2$ (Dashed line). Here, $\beta$ was determined by the fit of the average in Fig. 3A and B. The prefactor $B$ is plotted in Fig. 3C to accurately compare the effect of ATP depletion. Contrary to our expectations, no significant difference arose; ATP depletion did not remarkably alter $\langle|\tilde{f}_{\mathrm{A}}(\omega)|^2\rangle$ (Fig. 3C).

We also calculate the total strength of non-thermal fluctuations $Q(t_{\mathrm{e}})$ following Eq. (7), using well-defined and directly measured quantities. As shown in Fig. 3D, $Q(t_{\mathrm{e}})$ significantly decreased for ATP-depleted HeLa cells. Therefore, we believe that $Q(t_{\mathrm{e}})$ reasonably indicated the reduction of intracellular metabolic activities that occured with ATP depletion. Under the ATP depletion protocol, a minute quanatity of ATP still remains within the cells though it is decreased by more than an order of magnitude (13). In fact, cells are not viable under the absolute depletion of intracellular ATP. The decrease of $Q(t_{\mathrm{e}})$ by an order of magnitude in ATP-depleted cells is consistent to the decrease of ATP concentration shown in Fig. 3E.

We proceed to investigate the total velocity fluctuations $\langle|\tilde{v}(\omega)|^2\rangle$ measured with PMR (solid curves) and the thermal fluctuations $\langle|\tilde{v}_{\mathrm{th}}(\omega)|^2\rangle = 2\omega k_{\mathrm{B}}T\alpha''(\omega)$ estimated from AMR (red circles), as shown in Fig. 3F and G. At frequencies where the FDT violation was observed, $\langle|\tilde{v}(\omega)|^2\rangle$ was almost constant over frequency, *i.e.*, $\langle|\tilde{v}(\omega)|^2\rangle \sim \langle|\tilde{v}_{\mathrm{A}}(\omega)|^2\rangle \propto \omega^0$. This frequency dependency of the non-thermal fluctuation is apparently equal to that of simple diffusion; it is thus refered to as active diffusion (39, 49). On the other hand, the thermal fluctuations were sub-diffusive (also referred to as anomalous), *i.e.*, $\langle|\tilde{v}_{\mathrm{A}}(\omega)|^2\rangle \sim \omega^{\nu}$ with $0 < \nu < 2$, over the same frequency range where active diffusion was observed. This is due to the viscoelasticity of the glassy cytoplasm; fluid-like behavior may be observed at frequencies lower than the range of our MR experiment.

We evaluate the non-thermal velocity fluctuation $\langle|\tilde{v}_{\mathrm{A}}(\omega)|^2\rangle$ by quantifying the FDT violation using Eq. (13). From $\langle|\tilde{v}_{\mathrm{A}}(\omega)|^2\rangle$, we compute the mechanical power $\langle v_{\mathrm{A}}(t)\cdot f_{\mathrm{A}}(t)\rangle$ using the right-hand side of Eq. (12). In Fig. S1, $\langle v_{\mathrm{A}}(t)\cdot f_{\mathrm{A}}(t)\rangle$ appears to decrease in ATP-depleted cells. Though this trend aligns with our expectation, it is worth noting that the results may incorporate systematic errors for reasons we discuss in the supplementary. In addition to that $\langle|\tilde{f}_{\mathrm{A}}(\omega)|^2\rangle$ did not reflect metabolic activities, there is also the issue of uncertain nature of $f_{\mathrm{A}}(t)$ as we will elaborate on later. Therefore, we primarily address this quantity in the supplementary material, rather than in the main text.

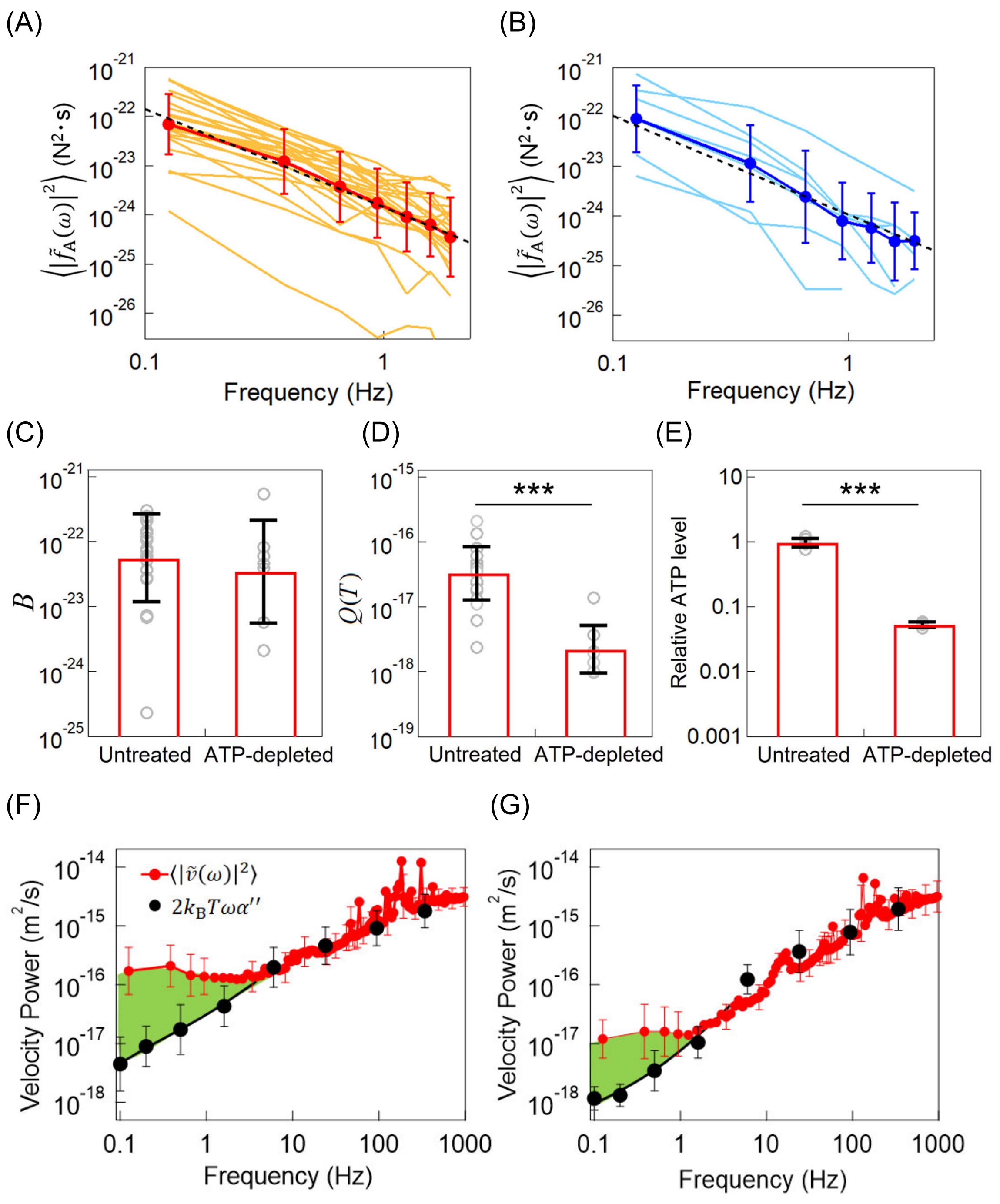

FIGURE 3 Power spectral density of non-thermal active force $\left\langle |\tilde{f}_{\mathrm{A}}(\omega)|^2 \right\rangle$ measured in (A) untreated and (B) ATP-depleted cells. Circles are the average and bars indicate the log-normal SD. Average values were fitted by $\left\langle |\tilde{f}_{\mathrm{A}}(\omega)|^2 \right\rangle = B\omega^{\beta}$ with $\beta = -2$, as shown by the dashed lines. (C) Magnitude of the non-thermal active force $B$ obtained by the fitting. No significant difference was observed, $P = 0.41$. (D) $Q(t_{\mathrm{e}})$ estimated by the area of green-colored region in Fig. 2 (A) and (B). The difference was significant, $P = 2.08 \times 10^{-4}$. (C, D) The Mann–Whitney U test was used to calculate the P values. ***: P<0.001. (E) The relative ATP concentration in cells, reproduced with modification from Ref. 13. (F, G) $\left\langle |\tilde{v}(\omega)|^2 \right\rangle$ (red circles) and $\left\langle |\tilde{v}_{\mathrm{th}}(\omega)|^2 \right\rangle = \omega 2 k_{\mathrm{B}} T \alpha''(\omega)$ (black circles) of probe particles embedded in (F) untreated and (G) ATP-depleted HeLa cells. In (A)(B) and (F)(G), curves are interpolations, and bars indicate the log-normal SD.

# DISCUSSION

## Microscopic and mesoscopic dynamics assumed in Langevin models

It was found that the active force spectrum $\langle |\tilde{f}_A(\omega)|^2 \rangle$ obtained based on the non-equilibrium Langevin model did not represent cellular metabolic activity, regardless of our expectations (30, 39, 50). To clarify the reason of this counter intuitive result, we first summarize the Langevin model for a mesoscopic probe particle (~ 1 μm) we measure, with attention to the assumptions and premises implicitly made to derive the model.

The equilibrium Langevin model was derived from a microscopic basis following Mori's projection operator formalism (51, 52). The derivation begins by examining the fluctuation of the mesoscopic probe particle [ $u(t)$ : observable] and determining the microscopic origin of the fluctuation. In an equilibrium medium, fluctuations are typically attributed to the thermal motions of solvent molecules, which occur at the scale of ~ Å (Fig. 4A). When applying a similar framework to living cells to obtain non-equilibrium Langevin model (40), it was implicitly assumed that the activities of mechano-enzymes were microscopic. The direct interaction with active enzymes then contributes to probe fluctuations, in addition to the thermal collision of small molecules (Fig. 4B). The fluctuating force $f_{th}$ or $f_A$ acting on the probe particle is obtained by integrating these microscopic interactions across the surface of the probe.

However, the motion of a mesoscopic probe does not directly reflect the microscopic dynamics, but rather more sensitively detects slow and long-wavelength fluctuation of surrounding medium. Therefore, the motion of each small molecule (or active enzyme) surrounding the mesoscopic probe is divided into mesoscopic flow and microscopic fluctuations. The former is equivalent to what appears in an athermal continuum in which a probe particle moves with $u(t)$ (Fig. 4A and B). Once the history of the probe motion [ $u(t)$ ] is given, the mesoscopic flow can be predicted by following the formalism of linear continuum mechanics.

Mesoscopic variables such as $u(t)$ , $f_{th}$ and $f_A$ fluctuate stochastically because microscopic dynamics exist behind the coarse-grained quantities. However, the complex processes mediating the microscopic and mesoscopic phenomena are not predictable owing to highly nonlinear microscopic interactions as depicted in Fig. 4C and D. Therefore, in the situation where the Langevin model is valid, the fluctuation of a mesoscopic probe does not measurably correlate with the microscopic dynamics of solvent molecules (or activities of mechano-enzymes). It is common to characterize the dynamics of a measured quantity by the statistical average of the correlations, *e.g.*, $\langle u(t)u(0) \rangle$ and $\langle f_{th}(t) f_{th}(0) \rangle$ . These correlations are related to the mesoscopic properties such as the response function $\alpha(t)$ and memory function $\gamma(t)$ via the FDT [Eq. (4) and (5)]. Thus, the fluctuation of a mesoscopic probe used with microrheology do not correlate with microscopic dynamics, and the characteristics of mesoscopic dynamics are determined by the mesoscopic properties of the system.

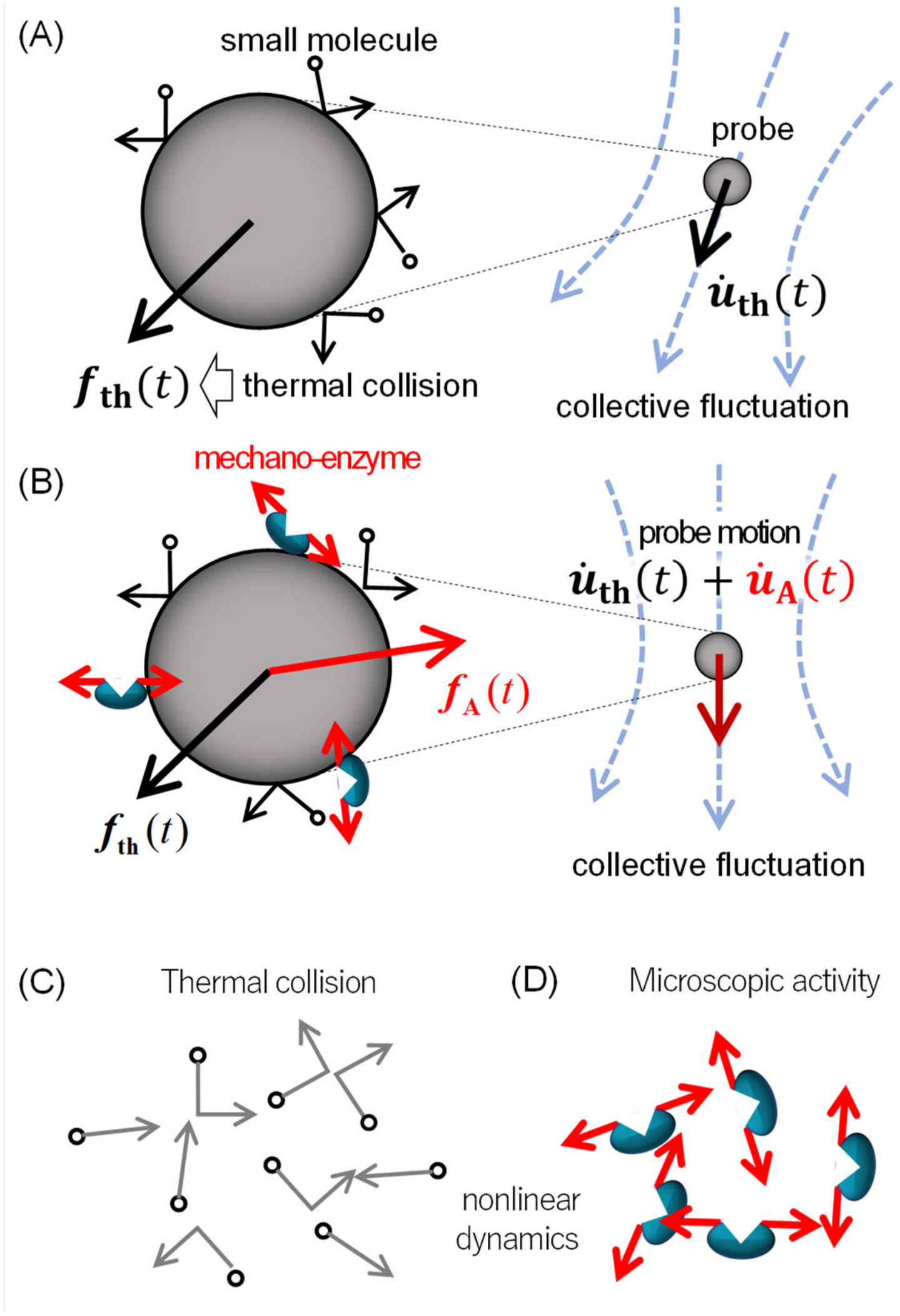

FIGURE 4 Microscopic and mesoscopic dynamics of fluctuations. (A) Conventional Langevin model at equilibrium. The thermal stochastic force $f_{th}(t)$ drives the movement of the probe particle [$\dot{u}_{th}(t)$] as given in Eq. (2). The colloidal particle accompanies the collective flow field in the medium (broken arrows). The microscopic dynamics of solvent molecules and the motion of the probe particle are mostly "separated", meaning that they are not directly correlated. (B) Langevin model extended to a non-equilibrium situation. Mechano-enzymes are assumed to act as microscopic objects, *i.e.*, their non-thermal activities are separated from the motion of the probe particle and the collective field in the surrounding medium. The thermal collisions of small solvent molecules and microscopic activities of mechano-enzymes are schematically shown in (C) and (D), respectively. Mesoscopic fluctuations in the medium do not directly correlate with the microscopic dynamics, owing to the nonlinear response of the microscopic processes.

## Non-microscopic dynamics of mechano-enzymes

While it was not explicitly stated in earlier studies, the non-equilibrium Langevin model is built on the assumption that the mechano-enzymes, acting as microscopic entities, directly interact with the probe particle (40). The force $f_{\mathrm{A}}(t)$ possesses a sound physical basis when this assumption holds true. Under such conditions, the microscopic activity of mechano-enzymes would not measurably correlate with the fluctuation of the colloidal particle or the mesoscopic fluctuations in the surrounding medium (Fig. 4B). A theoretical work (40), which introduced the non-equilibrium Langevin model, carefully refrained from asserting that $\langle|\tilde{f}_{\mathrm{A}}(\omega)|^2\rangle$ reflects the molecular dynamics of motor activity. In our experiments, mechano-enzymes are sterically repelled from the PEG-coated probe particles, leaving the meaning of $f_{\mathrm{A}}(t)$ in living cells ambiguous.

It has been shown, however, that the probe motion in the active medium directly reflects the dynamics of motor-generated forces (25-27, 30). Motor-generated forces directly induce mesoscopic flows in the surrounding medium (Fig. 5C and D). For instance, a myosin contraction in an F-actin network (Fig. 5A) leads to anomalous anti-parallel correlations among distant pairs of probe particles (25, 26), depending on the specific geometrical arrangements of the particle pairs and the (small number of) active mechano-enzymes in their vicinity. Kinesin or dynein motors transporting nm ~ μm cargos induce mesoscopic flows (active burst) in the surrounding medium (53) (Fig. 5B). With these mechanical perturbations, the stormy (chaotic or turbulent-like) dynamics are commonly observed in oocytes and embryos (54).

Although AMR does not probe the mechanical properties of sparse cytoskeletons, it does not contradict with the role of the motor-cytoskeleton complex in force transmission. During AMR, when an external force is applied to the PEG-coated probe particle, it barely experiences a restoring force from the sparse cytoskeletal networks. However, these networks, in conjunction with motor proteins, generate mesoscopic flows. In these larger-scale flows, it is reasonable to consider the sparse cytoskeletal network and the cytoplasm as a single unified continuum because the viscoelastic coupling between them makes decoupling difficult (25, 55). Other mechano-enzymes could also generate mesoscopic fluctuations while they change their shapes or make steps during catalytic turnover (18, 19, 56); dynamics with long-range correlation is one of characteristic features of crowded glassy systems. Besides, as we write in the introduction, “mechano-enzyme” includes submicrometer-sized organelle which is a complex of various enzymes.

In our experiments, the probe particle then moves by following the mesoscopic flow created by mechano-enzymes. The frictional force is minimal due to the small velocity difference between the probe and the medium. This condition is inconsistent with the requirements of the Langevin model, making the basis of $f_{\mathrm{A}}(t)$ defined in the non-equilibrium Langevin model elusive. We therefore model such a mechano-enzyme as a non-microscopic (mesoscopic) object which generates mesoscopic flow in the surrounding medium. Over the time scale where the FDT is violated, a mechano-enzyme is modelled as a force dipole $\kappa(t)$ since force balance is satisfied for slow over-damped motion of a small object in a highly viscous medium (25, 57). For simplicity, consider an active force dipole in a continuum. The displacement $\boldsymbol{u}_{\mathrm{A}}(t;\boldsymbol{r})$ at position $\boldsymbol{r}$ from the force dipole is given as shown in Fig. 5C by

$$\boldsymbol{u}_{\mathrm{A}}(t;\boldsymbol{r}) = \frac{3a\left[1-3(\hat{\boldsymbol{r}}\cdot\hat{\boldsymbol{n}})^2\right]\hat{\boldsymbol{r}}}{4r^2}\int_{-\infty}^{t}\alpha(t-t')\kappa(t')dt' . \tag{14}$$

The definition of parameters such as $\hat{\boldsymbol{r}}$ and $\hat{\boldsymbol{n}}$ in Eq. (14) are given in the caption of Fig. 5C. The response function $\alpha(t)$ is the same as that given in Eq. (2). A probe particle fluctuates by following this mesoscopic force-dipolar field, ensuring direct correlation between $\kappa(t)$ and $u_{\mathrm{A}}(t)$. By contrast, in the situation where the non-equilibrium Langevin model is relevant, a monopolar force field should emerge around the probe particle as shown in Fig. 4B. As the enzymes act in a microscopic manner, like solvent molecules colliding with the probe, they do not directly generate mesoscopic fluctuations (40).

**Not force but displacement is regulated during enzymatic reactions**

In the mesoscopic continuum model given in the previous section, the non-thermal displacement field $\boldsymbol{u}_{\mathrm{A}}(t;\boldsymbol{r})$ arises in response to a force dipole $\kappa(t)$. The model inherently assumes $\kappa(t)$ as the control parameter. However, the actual parameter regulated during enzymatic reactions are the conformation of the mechano-enzyme and resultant displacement fluctuation in the surrounding medium (56), not $\kappa(t)$. During a catalytic process, an enzyme sequentially transitions through various states, each linked to a unique conformation of the macromolecule. The shifts in atomic positions within the enzyme molecule during catalysis span in a range from less than Å to nm, depending on the enzyme and reaction involved (58, 59). For instance, motor proteins such as conventional kinesins and myosins take steps of 5 ~ 8 nm per ATP hydrolysis, irrespective of the mechanical load (friction/resilience) they face (17, 60). Therefore, mechano-enzymes must generate larger forces under higher mechanical loads, to attain the required conformational change or steps.

In our study, 'efficiency' describes the mechanical work done per ATP consumed to generate non-thermal fluctuations, a measure of the conversion efficiency from chemical to mechanical energy dependent on the mechanical load (16, 61). This deviates from the typical interpretation of 'efficiency' in motor protein research, which focuses on the distance covered per unit time. Under heavier loads, while the stepping rate and velocity of the motor decrease, the force exerted per step increases. As a result, the total work per ATP can increase under load, given that work is the product of force and distance. Near the stall force, motors display nearly equal forward and backward stepping rates. Even though this might seem futile at the macroscopic level because there's no net movement, these steps generate non-thermal mesoscopic fluctuations that we measure using PMR. In this context, ATP consumption is not wasted. We also note the possibility of futile ATP consumption, occurring likely under extremely high loads, where a motor hydrolyzes ATP but fails to make a step. The effect of such futile consumption on the mechanical perturbation of the surrounding medium remains an open question. Significantly, our ATP-depleted cells continued to produce non-thermal fluctuations, indicating the sustained activity of mechano-enzymes.

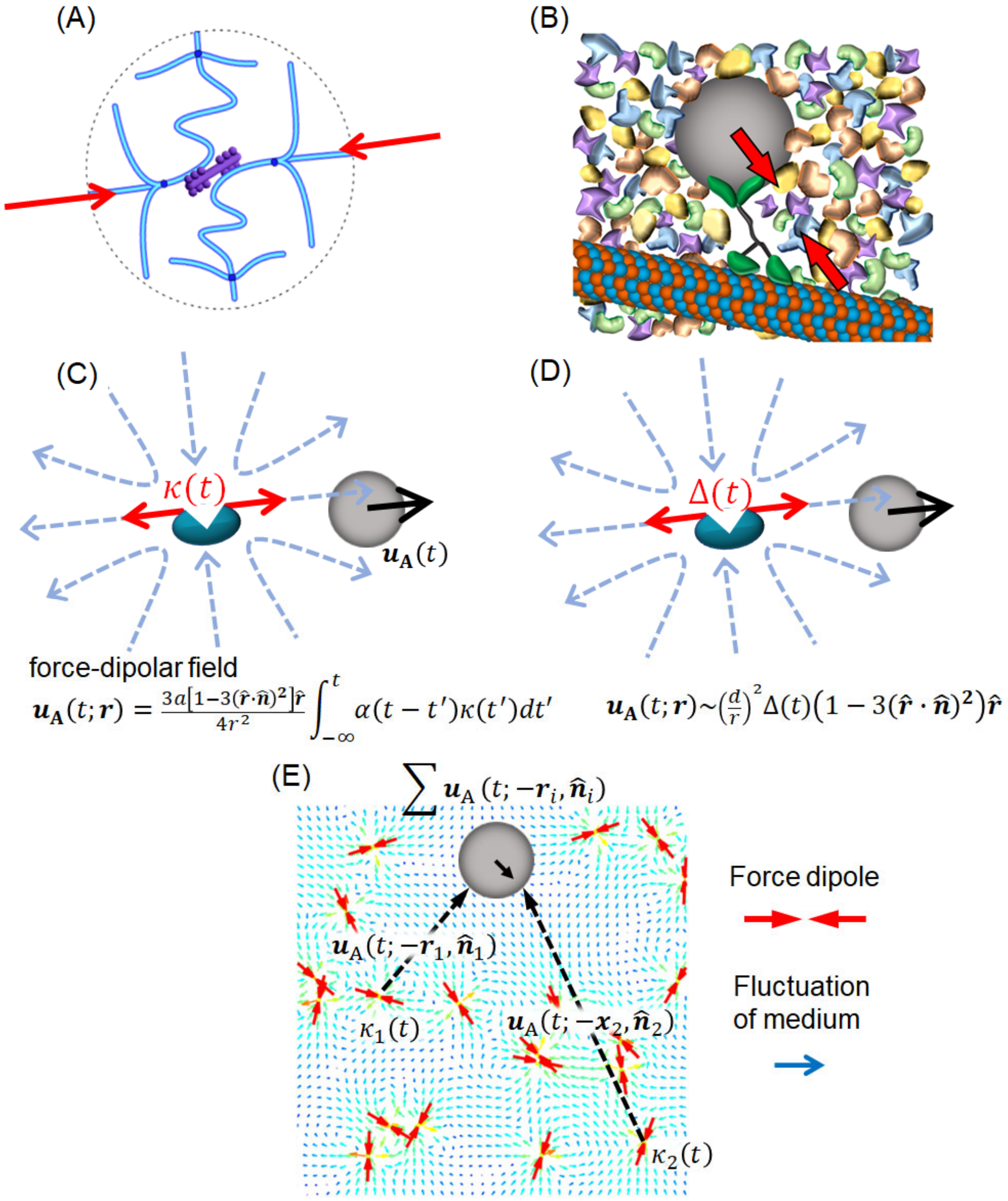

FIGURE 5 In living cells, mechano-enzymes generate mesoscopic fluctuation that directly correlates with their activity. The activity could be, *e.g.*, (A) acto-myosin contraction, (B) transportation of cargo in crowded environment, (C, D) the conformational change of mechano-enzymes. (C) An active mechano-enzyme modeled as force dipole $\kappa(t)$ and placed at $-\boldsymbol{r}$ from the probe particle induces a force-dipolar collective field $\boldsymbol{u}_A(t;\boldsymbol{r})$ which the probe particle follows. $\hat{\boldsymbol{n}}$ and $\hat{\boldsymbol{r}}$ are unit vectors orienting the direction of the dipolar force and $\boldsymbol{r}$, respectively. $\boldsymbol{u}_A(t;\boldsymbol{r})$ is delayed compared to $\kappa(t)$ owing to the viscoelasticity of the medium. (D) Response of the collective field $\boldsymbol{u}_A(t;\boldsymbol{r})$ to $\Delta(t)$ is instant over cellular length scales (~ μm) and at the time scale where FDT violation is observed (~ Hz). $\boldsymbol{u}_A(t)$ directly correlates with $\Delta(t)$ over mesoscopic length scales smaller than $\sqrt{|G|/\rho\omega^2}$ . (E) Fluctuation of the probe particle in a living cell is driven by a population of mechano-enzymes. Randomly distributed independent mechano-enzymes produce active fluctuations given by Eq. (A1).

Mechano-enzymes in living cells face mechanical loads that are tied to the viscoelasticity of the surrounding medium. In ATP-depleted cells, the force and mechanical work per ATP consumed tend to increase as $|G(\omega)|$ surpasses that of normal cells. This rise in efficiency in producing non-thermal fluctuations, resulting from increased viscoelasticity, clarifies why the force spectrum $\langle |\tilde{f}_{\mathrm{A}}(\omega)|^2 \rangle$ doesn't notably diminish in ATP-depleted cells, as illustrated in Fig. 3A and B.

**Displacement fluctuation is linked to metabolic activity**

In modeling the modulation of conformation instead of dipole strength, we represent the mechanical state (*e.g.* conformation or step) of an active mechano-enzyme as a length $\Delta(t)$, as illustrated in Fig. 6A. When $\Delta(t)$ changes, the displacement field in a medium of density $\rho$ rapidly responds over a length $\sim\sqrt{|G|/\rho\omega^2}$ in a time scale of $\sim 1/\omega$ (25, 62). In the cellular context, where the length $\sqrt{|G|/\rho\omega^2}$ typically exceeds ~ cm at frequencies showing the FDT violation ($\omega \le 1$ rad/s), an approximation akin to low Reynolds-number hydrodynamics applies to non-thermal fluctuations observed in cells. This is because the momentum transfer to the surrounding medium is fast compared with the slow viscoelastic response at each location in the medium. The active mechano-enzyme then induces a "displacement" field in the surrounding medium effectively instantly balancing with $\Delta(t)$:

$$\boldsymbol{u}_{\mathrm{A}}(t;\boldsymbol{r},\hat{\boldsymbol{n}}) \sim \left(\frac{d}{r}\right)^2 \Delta(t)\left(1-3(\hat{\boldsymbol{r}}\cdot\hat{\boldsymbol{n}})^2\right)\hat{\boldsymbol{r}}. \tag{15}$$

Note the consistency of Eq. (14) and Eq. (15). The difference is that the control parameter simply shifts from $\kappa(t)$ to $\Delta(t)$. Eq. (15) is particularly relevant in this study because $\Delta(t)$ is the variable controlled in cells and therefore better correlates with a cell's metabolic activity. In response to a given $\Delta(t)$, $\kappa(t)$ exhibits delay, *i.e.*, $\tilde{\kappa}(\omega) \sim 8\pi d^2 G(\omega)\tilde{\Delta}(\omega)$ or equivalently $\kappa(t) \sim 8\pi d^2 \int_{-\infty}^{t} G(t-t')\Delta(t')dt'$, and stress response at each location as well. On the other hand, $\boldsymbol{u}_{\mathrm{A}}(t;\boldsymbol{r},\hat{\boldsymbol{n}})$ directly correlates with $\Delta(t)$, without retardation and without depending on $G(\omega)$. The probe fluctuation is given by the sum of fluctuations produced by each force dipole positioned at $\boldsymbol{r}_i$ with orientation $\hat{\boldsymbol{n}}_i$ as $\boldsymbol{u}_{\mathrm{A}}(t) = \sum_{|\boldsymbol{r}_i|>a} \boldsymbol{u}_{\mathrm{A}}(t;-\boldsymbol{r}_i,\hat{\boldsymbol{n}}_i)$, as schematically shown in Fig. 5E. The PSD is derived in Appendix as

$$\langle |\tilde{u}_{\mathrm{A}}(\omega)|^2 \rangle = \frac{c\langle |\tilde{\kappa}(\omega)|^2 \rangle}{90\pi a|G|^2} = \frac{2cd^4}{9a}\langle |\tilde{\Delta}(\omega)|^2 \rangle. \tag{16}$$

where $c$ is the average number density of force dipoles. The activity of mechano-enzyme, as now embodied by $\Delta(t)$, is better probed by the non-thermal fluctuation of a probe particle $\langle|\tilde{u}_{\mathrm{A}}(\omega)|^2\rangle$ (25, 57) rather than by $\langle|\tilde{\kappa}(\omega)|^2\rangle$. The non-equilibrium Langevin model also leads to Eq. (16) if the PSD of the virtual force $f_{\mathrm{A}}$ was "formally" related with $\langle|\tilde{\Delta}(\omega)|^2\rangle$ such that $\langle|\tilde{f}_{\mathrm{A}}(\omega)|^2\rangle \propto |G|^2 \langle|\tilde{\Delta}(\omega)|^2\rangle$. It is clarified that $\langle|\tilde{f}_{\mathrm{A}}(\omega)|^2\rangle$ failed to reflect the metabolic activities in cells since the increase of $|G|$ cancelled the decrease of $\langle|\tilde{\Delta}(\omega)|^2\rangle$ according to the decrease of ATP.

Let us consider the PSD of non-thermal fluctuations, based on the mesoscopic continuum model using $\Delta(t)$ as a control parameter. During catalytic turnover, we assume that $\Delta(t)$ typically cycles between 0 and $\Delta_0$ over an average duration $\tau_{\mathrm{A}} = \tau_{\mathrm{b}} + \tau_{\mathrm{c}}$, where $\tau_{\mathrm{b}}$ and $\tau_{\mathrm{c}}$ are the times taken to make a transition to an excited state and then back to the original state. Also, we assume that each reaction takes place randomly with average frequency $\langle 1/\tau \rangle$, as shown in Fig. 6A. By employing $\tau_{\mathrm{A}} \sim \tau_{\mathrm{b}} > \tau_c$ for brevity, the frequency dependency of $\langle|\tilde{\Delta}(\omega)|^2\rangle$ is approximately obtained as shown in Fig. 6B

$$\langle|\tilde{\Delta}(\omega)|^2\rangle \sim \begin{cases} (\tau_{\mathrm{b}}\Delta_0)^2/\tau & (\omega < 1/\tau_{\mathrm{b}}) \\ (\Delta_0^2/\tau)\omega^{-2} & (1/\tau_{\mathrm{b}} \le \omega \le 1/\tau_{\mathrm{b}}) \\ (\Delta_0^2/\tau_c^2\tau)\omega^{-4} & (1/\tau_c < \omega) \end{cases}. \tag{17}$$

$Q(t_{\mathrm{e}})$ in Eq. (8) does not depend on the low-frequency cutoff at $\omega = 2\pi/t_{\mathrm{e}}$ because $t_e \gg \tau_{\mathrm{A}} \sim \tau_{\mathrm{b}}$ is satisfied in ordinary experimental conditions. The total power of probe displacements $Q$ [Eq. (7)] is then nearly equal to $Q(t_{\mathrm{e}}) \sim (cd^4/a)(\tau_{\mathrm{A}}\Delta_0^2/\tau)$, which we now know is proportional to the frequency of ATP hydrolysis $\langle 1/\tau \rangle$. Though $\tau_{\mathrm{A}}$ might slightly depends on ATP concentration, other parameters are supposed to be constant, which may explain why $Q(t_{\mathrm{e}})$ reflects the enzymatic activity in living cells.

$\langle|\tilde{\Delta}(\omega)|^2\rangle \propto \langle|\tilde{v}_{\mathrm{A}}(\omega)|^2\rangle$ is obtained as shown by the solid curve in Fig. 6C (57). In Fig. 6D, $\langle|\tilde{v}_{\mathrm{A}}(\omega)|^2\rangle$ in a tightly crosslinked active cytoskeleton (*in vitro* actin/myosin gel) is shown, which was calculated from the data presented in prior articles (25, 26). $\langle|\tilde{v}_{\mathrm{A}}(\omega)|^2\rangle$ showed a plateau around 1 ~ 10 Hz but is decreased at both higher and lower frequencies. The frequency dependency was similar to the green solid curve in Fig. 6C, indicating that Eq. (A3) is valid for the active cytoskeletons. Similar results are also obtained in an active medium whose meso-scale internal structure is maintained regardless of active fluctuations, *e.g.,* in the lamellipodia in living cells (20).

By contrast, $\langle|\tilde{v}_{\mathrm{A}}(\omega)|^2\rangle$ in living cytoplasm (Fig. 3E and F) does not conform to the prediction; the trend of the solid curve displayed in Fig. 6C differs qualitatively from the active diffusion we observed in Fig. 3E and F. In the coming sections, we will extend this discussion to a glassy medium, typically subject to structural relaxations under mechanical perturbations, and show that $Q(t_{\mathrm{e}})$ continues to be a measure of metabolic activities in the living cytoplasm. Prior to that, we must examine the mechanical situation of cells, which we propose to be in a critical jamming state. Subsequently, drawing upon the assumption, we integrate the most recent understanding of how melting occurs in dense disordered systems into our mesoscopic continuum model.

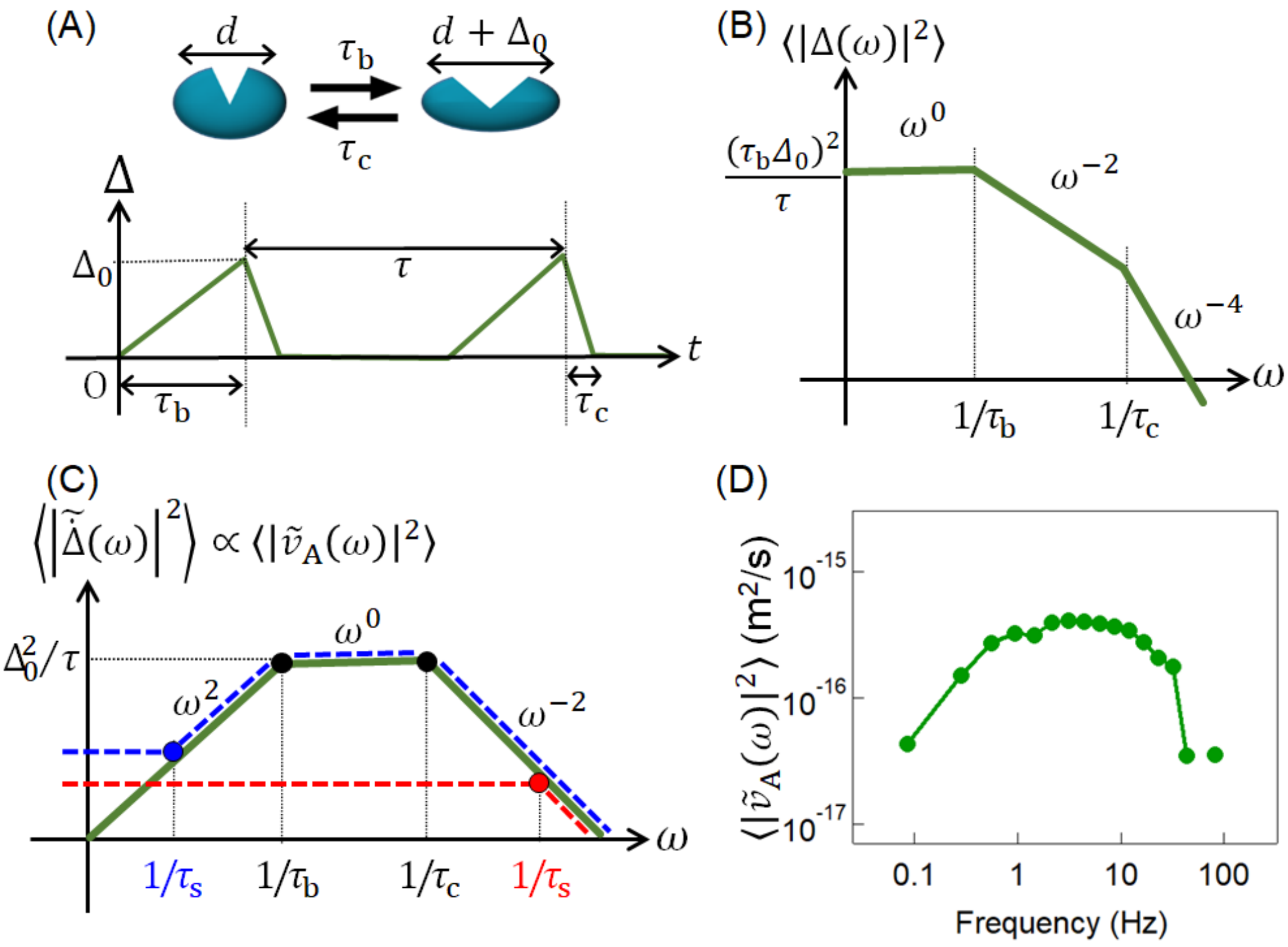

FIGURE 6 (A) Schematic of length $\Delta(t)$ representing action of a mechano-enzyme, *e.g.*, the conformation of a mechano-enzyme with a typical resting size *d*. $\Delta(t)$ typically cycles in $\tau_A = \tau_b + \tau_c$ during each enzymatic reaction which takes place with average frequency $\langle 1/\tau \rangle$. Note that the time interval $\tau$ between the successive enzymatic reactions is a stochastic variable, and $\langle 1/\tau \rangle$ represents enzymatic activity. The reaction events occur randomly, and the orientation $\hat{\boldsymbol{n}}$ and the position $-\boldsymbol{r}$ of the enzyme from the probe particle is altered stochastically at every catalytic reaction. (B, C) The schematic of the PSD for (B) $\langle |\Delta(\omega)|^2 \rangle$ and (C) $\langle |\tilde{\Delta}(\omega)|^2 \rangle \propto \langle |\tilde{v}_A(\omega)|^2 \rangle$ which were estimated from $\Delta(t)$ given in (A), by following the discussion in Ref. 57. The green solid curve is the case when there is no yielding. We will discuss later concerning the case when mechano-enzymes drive yielding over a time scale of $\tau_s$ (see Figs. 7A and 8D). The spectrum will then be altered for $\omega < 1/\tau_s$ as shown by the blue and red broken curves, depending on the corresponding $\tau_s$. (D) $\langle |\tilde{v}_A(\omega)|^2 \rangle$ of 2*a* = 5 μm colloidal particles embedded in a tightly crosslinked active cytoskeletons (actin/myosin gel), calculated from the data given in Fig. 6 of Ref. 26. $\langle |\tilde{v}_A(\omega)|^2 \rangle$ shows a plateau around 1 ~ 10 Hz but decreases at both higher and lower frequencies. The frequency dependency was similar to the green solid curve in (C), indicating that Eq. (A3) is valid. In contrast to the living cytoplasm, active diffusion was not observed because the tightly crosslinked actin gel hardly experiences structural relaxations.

## The active glassy cytoplasm lies at the transition between solid and liquid

Recently, it was found that a cytoplasm devoid of a cytoskeleton and metabolic activities possesses glass-forming capabilities (11, 12). A cytoplasm becomes glassy when the concentration of polymeric components increased and/or metabolic activities are inhibited (9-12). For instance, the viscosity of cytoplasmic extracts exhibits super-exponential increase as a function of solid content, reaching immeasurable levels at physiologically relevant concentrations (5). In living cells, multitudes of mechano-enzymes perturb the crowded cytoplasm by utilizing the energy released by ATP hydrolysis. The glassy cytoplasm is then fluidized; vigorous active fluctuations, as evidenced by the FDT violation, trigger structural relaxations that would not occur solely with thermal activations (12, 20). Therefore, the living cytoplasm represents an "active glass" (63) or "dense active matter" (64), which are currently the subject of intensive theoretical investigation. Previous theoretical studies have predicted that the dense system becomes increasingly fluidic as the active driving force increases (65, 66).

However, the fluidization of the living cytoplasm does not fully align with these theoretical predictions, but halts midway towards complete fluidization. Even though the elastic plateau $G_0$ in ATP-depleted cells vanishes in normal cells, $G(\omega) \propto (-i\omega)^{1/2}$ is observed instead of the anticipated fluid response $G(\omega) \propto -i\omega\eta$ with a constant viscosity $\eta$. This indicates that normal cells poise at the threshold of solid and liquid states, without progressing towards a more fluid state, contrary to theoretical predictions. This phenomenon is consistent to the manner in which mechano-enzymes operate in a highly viscoelastic medium. As discussed in the preceding section, the mechanical energy (or force) generated by the mechano-enzymes is efficiently transmitted to the surrounding medium when the mechanical loads [viscoelasticities $G(\omega)$] are higher (16, 61). However, when the elastic response of the cytoplasm is lost owing to metabolic activities, the mechano-enzymes does not efficiently fluidize the surrounding medium. Such regulation of perturbations in an active medium has not been accounted for in previous theories of active glass/dense active matter.

## The critical jamming rheology of active glassy cytoplasm

The power-law rheology $G(\omega) \propto (-i\omega)^{1/2}$ observed in living cells was not anticipated in the theories for active glass, or dense active matter. However, it has been predicted for athermal disordered systems close to critical jamming transition (13). The jamming transition at $T = 0$ is not just a crossover between fluid and solid states; they are critical on the verge of mechanical stability, with the number of contacts between crowding objects satisfying isostaticity (6 in 3 dimensions). This state of marginal stability can sustain infinitesimal stress (like a solid), but any additional stress induces a rearrangement of constituents and a transition to a new configuration (like a liquid). The marginal stability at such critical juncture results in an indefinitely extending distribution of relaxation times for anomalous non-affine modes, giving rise to the power-law rheology $G(\omega) \propto (-i\omega)^{1/2}$ (67, 68).

However, in dense suspensions of inert particles such as emulsion, hydrogel and foam, this critical situation is not typically observed; $G(\omega) = G_0 + A(-i\omega)^{1/2}$ was observed both below and above such geometric points (68-70). This is because an elastic plateau appears in $G'(\omega)$ due to thermal collisions between particles, which induce more effective contacts and contribute to the entropic elasticity, even at the geometrical condition that satisfies isostaticity. This scenario may apply to ATP-depleted cells showing $G(\omega) = G_0 + A(-i\omega)^{1/2}$ because they are almost at equilibrium as demonstrated by smaller deviation from the FDT. On the other hand, when dense suspensions are mechanically activated by *e.g.*, shear flow, the static elasticity $G_0$ disappears, and a frequency dependency similar to metabolically active cells [ $G(\omega) \propto (-i\omega)^{1/2}$ ] emerges (71, 72). While the subtle thermal fluctuations obscure the critical jamming state by stiffening the system, non-thermal fluctuations fluidize the system, placing it on the edge of the critical juncture. During the fluidization, some contacts between crowded objects destabilizes, leading to a decrease in the effective number of contacts towards the isostatic point. Even though intracellular environments are subjected to uncorrelated localized perturbations, it is expected that dense active matter undergo similar fluidization under macroscopic shear (73). Although specifics warrant further theoretical exploration, such as the effect of unique spatio-temporal dynamics and statistics of thermal and non-thermal fluctuations on fluidization, we thus presume that living cytoplasm operates under condition similar to critical jamming from a mechanical standpoint.

**Sub-diffusive thermal fluctuation and active diffusion in the living cytoplasm**

In a material expressing the power-law rheology $G(\omega) \propto (-i\omega)^{\beta}$ with $\beta \sim 0.5$ , the thermal fluctuation is sub-diffusive such that $\langle |\tilde{v}_{\mathrm{th}}(\omega)|^2 \rangle = 2\omega k_{\mathrm{B}} T \alpha'' \propto \omega^{1-\beta}$ (Fig. 3F). While $\langle |\tilde{v}_{\mathrm{th}}(\omega)|^2 \rangle \propto \omega^0$ was expected for pure thermal diffusion in simple liquids (Brownian motion), both the ATP-depleted and ordinary cytoplasm did not show this behavior. Apparently, the terminal relaxation time ($\alpha$-relaxation time $\tau_\alpha$) is longer than the range of frequencies available in our AMR experiments ( $\omega > 1/\tau_\alpha$ ). On the other hand, non-thermal fluctuation $\langle |\tilde{v}_{\mathrm{A}}(\omega)|^2 \rangle$ is not constrained by the FDT, and therefore their frequency dependence was not *a priori* known. Nevertheless, apparently diffusion-like behaviors, $\langle |\tilde{v}_{\mathrm{A}}(\omega)|^2 \rangle = \omega^2 \langle |\tilde{u}_{\mathrm{A}}(\omega)|^2 \rangle \propto \omega^0$, was observed at low frequencies for both normal and ATP-depleted HeLa cells as seen in Fig. 3 E and F. This frequency dependency is the same as that of thermal Brownian motion (74), and indicates that the auto-correlation is lacking in the active velocity of the probe particle, *i.e.*, $\langle v_{\mathrm{A}}(t'+t) v_{\mathrm{A}}(t') \rangle \propto \delta(t)$ . We define the apparent diffusion constant $D_{\mathrm{A}}$ describing the intensity of the active diffusion as $2D_{\mathrm{A}} \equiv \lim_{\omega \to 0} \langle |\tilde{v}_{\mathrm{A}}(\omega)|^2 \rangle$ for later discussion. Similar behavior, referred to as active diffusion (16, 20, 39, 49), has been ubiquitously observed in living cells and other motor-driven systems (16).

**Active diffusion as a result of non-microscopic yielding**

The occurrence of active diffusion is not immediately explained by either the non-equilibrium Langevin model or the mesoscopic continuum model. To further elucidate this for the mesoscopic continuum, let's examine the scenario where $\Delta(t)$ cycles over an average duration $\tau_{\mathrm{A}}$ as depicted in Fig. 6A. For reasons given in the caption, we disregard any correlation between metabolic reaction events. For $t' \gg \tau_{\mathrm{A}}$, we then derive $\langle \Delta(t+t')\Delta(t) \rangle \sim 0$, which indicates $\langle |\tilde{\Delta}(\omega)|^2 \rangle \propto \omega^0$ for $\omega \ll 1/\tau_{\mathrm{A}}$. By applying Eq. (A3), the continuum model provides $\langle |\tilde{u}_{\mathrm{A}}(\omega)|^2 \rangle \propto \omega^0$, in contrast with the active diffusion [$\langle |\tilde{v}_{\mathrm{A}}(\omega)|^2 \rangle \propto \omega^0$] (75).

The emergence of active diffusion indicates that active velocity $v_{\mathrm{A}}(t)$ undergoes memory loss owing to structural relaxations (yielding). When a material deformed beyond its elastic limit yields and transitions to plastic (permanent) deformation. Behind this macroscopic yielding exist microscopic rearrangements of molecules. To investigate the microscopic mechanism of the glass 'melting' processes, which is fundamentally driven by these local yields, 'yielding' also refers to the local structural rearrangements that occur in a viscoelastic medium. It is then plausible that bio-macromolecules confined in a crowded cytoplasm would alter their configuration (termed as an inherent structure in glass study) when they are perturbed by the forces generated by active mechano-enzymes. We refer to this phenomenon as 'active structural relaxation', or 'active yielding'. However, if these structural relaxations occur on a microscopic scale as illustrated in Fig. 7A, the mesoscopic continuum model again fails to predict active diffusion. Even if microscopic yielding takes place, it merely modifies the medium's continuum mechanical property, $G(\omega)$. Even if the medium evolves into purely viscous state, *i.e.*, $G(\omega) \sim i\omega\eta$ with $\eta$ representing viscosity, the probe particle's response $u_{\mathrm{A}}(t)$ to the mesoscopic activity $\Delta(t)$ is still governed by Eq. (15). Eqs. (14) (15), (A3) continue to describe the non-thermal fluctuations driven by the mechano-enzymes. With the same reason as in the previous paragraph, the mesoscopic continuum model does not result in active diffusion, even with microscopic yielding. Non-microscopic yielding that accompanies mesoscopic fluctuations as portrayed in Fig. 7B is necessary to explain active diffusion.

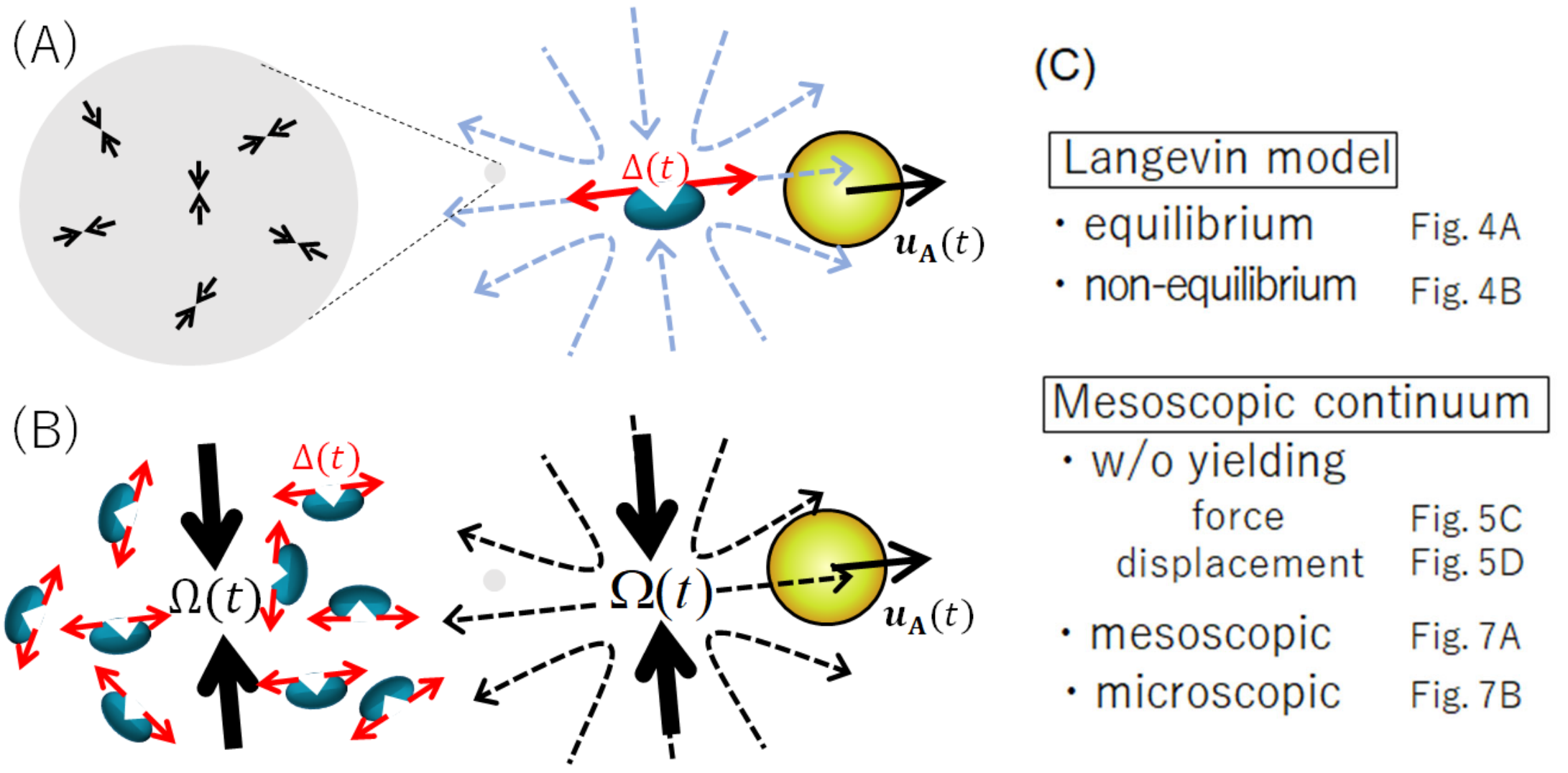

FIGURE 7 (A) Microscopic yields (black dipolar arrows in left) do not directly generate mesoscopic flows while they govern mesoscopic viscoelasticity of medium. Correlation between activity of mechano-enzymes $\Delta(t)$ and mesoscopic flow $\boldsymbol{u}_A(t)$ is maintained in this case, *i.e.*, active diffusion should not occur. (B) Activity of crowded mechano-enzymes $\Delta(t)$ induces a mesoscopic yield represented by $\Omega(t)$ that generates a mesoscopic flow in the surrounding medium. The mesoscopic flows generated by $\Delta(t)$ and $\Omega(t)$ lose their direct correlation because a plastic yielding is highly nonlinear process, leading to active diffusion. (C) Summary of models.

**mesoscopic yielding gives rise to force-dipolar fluctuation and active diffusion**

According to the most recent understanding on the melting of dense disordered systems, long-range meso-scale fluctuations appear upon the onset of structural relaxation (76, 77). Notably, yielding leaves a force-dipolar displacement field in the surrounding medium. In Fig. 8A, we show a typical example of fluctuations of densely confined particles. Brownian dynamics simulation was performed under conditions described in Supplementary (78-80). Arrows indicate thermal displacements that occurred in ~ 0.3 $\tau_\alpha$. Particles are rearranged at several yield spots in the overall simulation area. Outside the yields, smooth continuous displacements that do not alter the neighboring pair of particles appear. Let us focus on the square area bordered by the broken line in Fig. 8A centered by the blue circle. When the radial components of the displacements in the area are extracted, a quadrupole symmetric pattern typical of a force dipole is found as shown in Fig. 8B. A similar pattern was observed also in a dense colloidal system where each colloid was independently driven far from equilibrium (81).

When individual colloidal particles are driven independently by either thermal (78) or non-thermal noise (65, 81), fluctuations that are not associated with yielding are also prevalent, obscuring the force dipolar fluctuations. It was therefore essential to isolate the radial distribution by removing the overall drift of the area. In glassy systems under a quasi-static shear at zero temperature, the elementary processes of structural relaxation can be unequivocally discerned. An example of such simulation results is shown in Fig. 8C. The quadrupole-type plastic displacement field, characteristic of those created by a force dipole was induced by a mesoscale structural relaxation (80, 82). The quadrupolar symmetric pattern was even more apparent without manipulating the original simulation results (Fig. 8C). Though perturbations in dense active matter (active glass), such as cells, are uncorrelated and localized, these systems melt essentially in a similar manner to those subjected to shear (73).

These mesoscopic fluctuation in a glass is consistent with an Eshelby field (76). As detailed in the classical theory (83), when a plastic deformation is implemented in elastic continuum, the lowest-order is the force dipole that cause quadrupolar symmetric displacements. This is because monopole is excluded by the internal force balance. Therefore, long-range mesoscopic flow which occur due to yielding in continuum is approximated by the collection of force dipolar fields with quadrupolar symmetry. In a very recent study, it was shown that thermally excited Eshelby-type quadrupolar fields underlies the important characteristics of glassy systems such as the dynamic heterogeneity (84). Even overcooled liquids which is somewhat fluid undergo structural relaxation via the accumulation of Eshelby-like plastic deformation sub-processes given by Eq. (14). Similar to typical glassy systems, the fluctuations observed in cells and active cytoplasm are non-Gaussian owing to the heterogeneous dynamics (11, 27, 85, 86). It is then plausible that similar mesoscopic relaxation plays an important role for non-thermal fluctuations in living cells.

Importantly, the yielding in a glassy material will permanently leave a force-dipolar displacement field in the surrounding medium. Then, let $\Omega(t)$ [instead of $\Delta(t)$ ] represent the deformation of the medium at a yield spot. As shown in Fig. 7, a mesoscopic relaxation event represented by $\Omega(t)$ generates a mesoscopic fluctuation (Fig. 7B) whereas microscopic yields do not (Fig. 7A). In the same way that diffusive fluctuations are induced by the random occurrence of such structural relaxations in space, active diffusion occurs when the glassy cytoplasm experiences mesoscopic structural relaxations (yielding) when mechanically driven by active mechano-enzymes. After a mesoscopic relaxation is induced by enzymatic activity, its trace is left as a permanent deformation in the surrounding media. Once yielding occurs, the probe particles both close to and distant from the event move in correlation with it, and do not return. Since fluctuations occur independently and randomly at every yielding event, the active fluctuation $u_{\mathrm{A}}(t)$ exhibits a diffusion-like frequency dependence (*i.e.* $\langle |\tilde{u}_{\mathrm{A}}(\omega)|^2 \rangle \propto \omega^{-2}$, $\langle |\tilde{v}_{\mathrm{A}}(\omega)|^2 \rangle \propto \omega^0$), which will dominate over the non-yield component $\langle |\tilde{v}_{\mathrm{A}}(\omega)|^2 \rangle \propto \omega^2$ (Fig. 6C) at long time scales.

**Qualitative model that links active diffusion to metabolic activity**

It is possible to incorporate such mesoscopic relaxation into the mesoscopic continuum model. The crucial observation is that mesoscopic structural relaxation occurs as if a permanent force dipole appears at the hot spot of the yielding event (76). Fig. 8A and B show the displacements around the local structural relaxation in a glassy system simulated *in silico*. The far-field displacements shown in Fig. 8A are the two-dimensional version of Eq. (15), *i.e.*, the force dipolar field. $\Omega(t)$ [instead of $\Delta(t)$ ] was defined as the deformation of the medium at a yield spot; each yielding event leaves permanent displacements induced by a local activity represented by a length $\Omega_0$, as shown in Fig. 8D. Because the orientation of the yield $\hat{\boldsymbol{n}}$ is random, let $\Omega(t)$ take positive or negative steps randomly with a frequency $1/\tau_{\alpha}$. $\Omega(t)$ then performs a one-dimensional random walk. Assuming that it takes $\tau_{\mathrm{s}}$ on average for each step to occur (see Fig. 8D), we have

$$\left\langle \left|\tilde{\Omega}(\omega)\right|^2 \right\rangle \sim \frac{2\Omega_0^2}{\tau_{\alpha}} \frac{1}{1+(\omega\tau_{\mathrm{s}})^2} . \tag{18}$$

If we let *d* be the size of the yield spot, and $c \sim 1/d^3$, Eq. (A3) is rewritten as

$$\langle |\tilde{v}_{\mathrm{A}}(\omega)|^2 \rangle \sim \frac{2d}{9a} \left\langle \left|\tilde{\Omega}(\omega)\right|^2 \right\rangle . \tag{19}$$

Incorporating Eq. (18) to this formula, we have active diffusion $\langle |\tilde{v}_{\mathrm{A}}(\omega)|^2 \rangle \sim d\Omega_0^2/(a\tau_{\alpha})$ for $\omega < 1/\tau_{\mathrm{s}}$ and super-diffusion for $\omega > 1/\tau_{\mathrm{s}}$, as schematically shown in Fig. 8D. Note that super-diffusion was hidden in thermal fluctuations in this study, but it was observed in prior studies in tightly crosslinked active cytoskeletons *in vivo* and *in vitro* (20, 26). For normal cells, we have $\lim_{\omega\to 0} \langle |\tilde{v}_{\mathrm{A}}(\omega)|^2 \rangle = 2D_{\mathrm{A}} \sim 10^{-16}$ m$^2$/s as shown in Fig. 3F. By estimating $d \sim \Omega_0 \sim 10^{-7}$ m from the mechanical inhomogeneity in cells (11), we have $\tau_{\alpha} \sim 10$ s which seems to be consistent to our recent experiments (data not shown).

In the continuum model with sparse mesoscopic yielding, active diffusion arises in $\omega < 1/\tau_{\mathrm{s}}$. On the other hand, the medium should fluidize for $\omega < 1/\tau_{\alpha}$ because $\tau_{\alpha}$ is nothing but $\alpha$-relaxation time. Because the mesoscopic flow is long ranged, the fluctuation induced by an yield reaches far beyond the local yielding spot. The non-thermal probe fluctuation may be measurably driven when only a single structural relaxation was activated at a location distant from the probe particle. On the other hand, note that the residual stress in a medium is released mostly at the local yielding spot. To achieve a macroscopic structural relaxation, everywhere in the specimen must directly experience the local yielding event, meaning $\tau_{\alpha} \geq \tau_{\mathrm{s}}$. In this study, the terminal relaxation in the living cytoplasm seemed too slow to observe with our MR technique whereas active diffusion was commonly observed. Our experimental results indicate that $\tau_{\alpha} \gg \tau_{\mathrm{s}}$ in accordance with a model where rare and sparse yields induce mesoscopic fluctuations. Because of this consistency, it is reasonable to expect that a purely fluid response [ $G(\omega) = -i\omega\eta$ ] will be observed if we could extend the frequencies available with AMR towards smaller $\omega$.

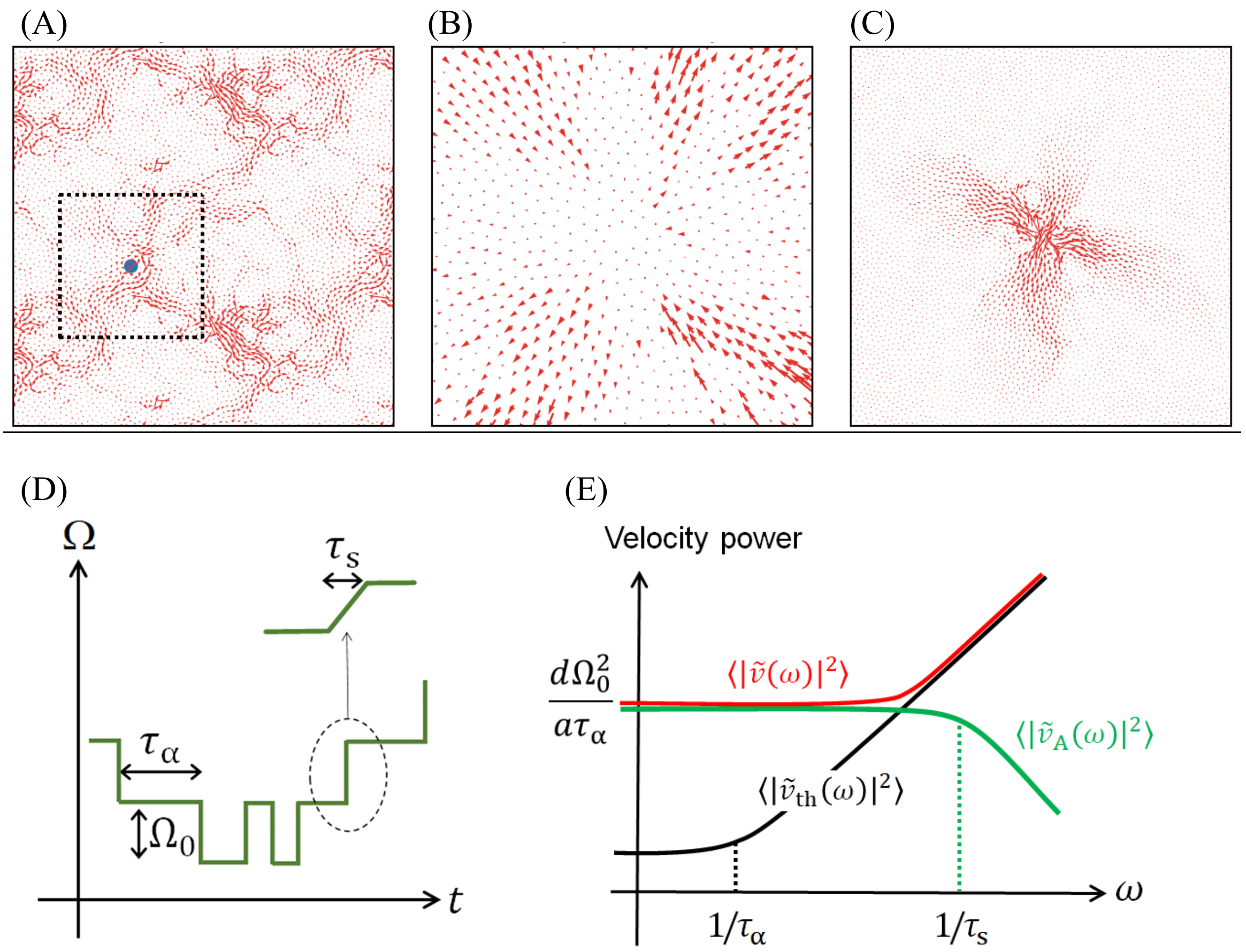

FIGURE 8 (A) Displacement field observed in a thermal glassy system. Displacements of particles during $\sim 0.3\ \tau_\alpha$ are shown. The data is taken under the same conditions given in ref. 79. (B) A close-up image of the subsystem from (A) marked by the dashed square. Only the radial component of displacements is shown. The center of the radial coordinate is chosen to be the blue point in (A). (C) Displacements during a single yield in a zero-temperature system which was subjected to quasistatic shear. The data is from ref. 81. Structural relaxation in a glassy system leaves permanent displacements similar to the force-dipolar field given in Eq. (16). See Supplementary for a detailed description of the numerical setups for panels (A-C). (D) Schematic for typical time evolution of $\Omega(t)$ representing the mesoscopic yields. We consider here the history of yields that occur at a certain yield location encircled in Fig. 7A and B. Although the orientation of the force dipole representing each yield is randomly distributed in 3D, we describe it in 1-dimensional form for brevity. (E) Schematic representation of the typical velocity power spectrum of living cytoplasm, given in Fig. 3E and F. Red and black curves indicate the power spectral density of the total and thermal velocity fluctuations, respectively. The green curve indicates the non-thermal fluctuations, obtained as the difference between the total and thermal fluctuations. Active diffusion is observed for $\omega < 1/\tau_s$ . On the other hand, we expect that the living cytoplasm is fluid-like for $\omega < 1/\tau_s$ though the frequencies are out of the range measurable with our MR technique.

We can now discuss the significance of the active diffusion constant. Consider the cytoplasm as a homogeneous continuum with mechanical properties designated by $G(\omega)$. Active mechano-enzymes trigger local yields with an average frequency $1/\tau_\alpha$. These yields could occur at the same location as active mechano-enzymes, being directly driven by the specific mechano-enzyme. However, theoretical studies have shown that yield spots can also manifest even where there are no active particles, which is a distinctive characteristic of crowded systems (65). Even though yielding and the activity of mechano-enzymes may not be explicitly correlated, we contend that their occurrence rates are, on average, related. Note that both $\Delta(t)$ and the active fluctuations incited by $\Delta(t)$ are slow variables, given the mesoscopic nature of mechano-enzymes. When the system locally relaxes in response to the slow changes of $\Delta(t)$, the mesoscopic dynamics is quasi-static and therefore can be normalized with the rate of the enzymatic reactions as a reference. Yielding events should then occur approximately at a frequency proportional to the average frequency of metabolic turnover $1/\tau$ of the enzymatic reactions, *i.e.*, $1/\tau_\alpha \propto 1/\tau$. An active diffusion constant is then obtained as

$$D_{\mathrm{A}} = \tfrac{1}{2}\lim_{\omega\to 0}\left\langle |\tilde{v}_{\mathrm{A}}(\omega)|^2 \right\rangle \sim \frac{2d\Omega_0^2}{9a\tau_\alpha} \propto 1/\tau . \tag{20}$$

We expect that the relation $\lim_{\omega\to 0}\left\langle |\tilde{v}_{\mathrm{A}}(\omega)|^2 \right\rangle \propto 1/\tau$ is robust since *d, a* and $\Omega_0$ do not greatly depend on activity in the glassy systems (87, 88). $2D_{\mathrm{A}} = \lim_{\omega\to 0}\left\langle |\tilde{v}_{\mathrm{A}}(\omega)|^2 \right\rangle$ then represents the frequency of ATP hydrolysis per unit period of time, *i.e.*, the active metabolism in the living cytoplasm.

When active diffusion is observed, we have $Q(t_{\mathrm{e}})/t_{\mathrm{e}} \sim 2D_{\mathrm{A}}$ by substituting $\left\langle |\tilde{u}_{\mathrm{A}}(\omega)|^2 \right\rangle = 2D_{\mathrm{A}}/\omega^2$ into Eq. (8). Both $Q(t_{\mathrm{e}})$ and $D_{\mathrm{A}}$ therefore represent metabolic activity in living cells if the experimental period of time $t_{\mathrm{e}}$ is fixed as we have done in our experiments. $D_{\mathrm{A}}$ can be defined only when the active diffusion is clearly observed with MR experiments. If the relaxation process is slow, $\tau_{\mathrm{s}}$ could be too large to measure active diffusion from MR experiments. Even if structural relaxation did not occur, for instance, in stably crosslinked active gels, $Q(t_{\mathrm{e}})$ will still correlate with metabolic activities, as we discussed in the paragraph including Eq. (A3). In any case, with or without active diffusion, $Q(t_{\mathrm{e}})$ may be a good measure for metabolic activities in living cells. In reality, the fluctuations due to mesoscopic yields and those directly driven by the mechano-enzymes coexist. When both are independent, the total non-thermal fluctuation would be simply obtained by the sum of Eq. (A3) and (19). On the other hand, when a mechano-enzyme directly induces yields at their respective locations, the frequency dependency of $\left\langle |\tilde{\Delta}(\omega)|^2 \right\rangle \propto \left\langle |\tilde{v}_{\mathrm{A}}(\omega)|^2 \right\rangle$ will be altered for $\omega < 1/\tau_{\mathrm{s}}$ as shown by the broken lines in Fig. 6 (C). Although the discussion above is an approximation, the results qualitatively explain the non-thermal fluctuations in various active soft materials as well as cells, as we will show in forthcoming papers.

In this study, non-thermal fluctuations in living cells were discussed by focusing on the mesoscopic dynamics because the fluctuations observed with MR seem to directly correlate with the activity of mechano-enzymes. This does not necessarily mean that the mechano-enzymes do not generate microscopic fluctuations. Because the probe particles used in MR experiments were mesoscopic, their movement predominantly reflects long-wavelength fluctuations. Though "mechano-enzyme" in this study includes broad entities of active agents in cells, the size of an enzyme molecule is typically several nm while the length scale of dynamic heterogeneity in living cells has been estimated to be 10 times greater (11). In the future, by conducting AMR with smaller beads (10 ~ 100 nm) in living cells, or by creating a model dense active matter (active glass) made of micrometer-sized active colloids, we may be able to investigate the microscopic dynamics of non-thermal fluctuations. When the dynamics of the probe particle reflect microscopic yielding more sensitively than a mesoscopic relaxation, active diffusion and thermal diffusion can take place at closer time scales, *i.e.*, $\tau_s \lesssim \tau_\alpha$, which may further support the notion that the non-thermal fluctuations observed in cells in this study were mesoscopic.

## CONCLUSION

In this study, we investigated the active mechanics of HeLa cells in a confluent epithelial monolayer. Probe particle movements, *i.e.*, the spontaneous fluctuation and the response to the applied force, were observed using the feedback-tracking microrheology (MR) technique (20, 57). Our results show that HeLa cells with ordinary metabolic activity violate the fluctuation-dissipation theorem (FDT) at low frequencies (< 10 Hz). The intracellular shear viscoelastic modulus showed a frequency dependence $G(\omega) \propto (-i\omega)^{1/2}$ which is typical for a glassy cytoplasm. Notably, when metabolic activity is decreased by ATP depletion, we observed a clear impact on the mechanical properties of the cell interior; slow non-thermal fluctuations were decreased while $G(\omega)$ exhibited an elastic plateau at low frequencies. These observations are consistent with recent studies which suggested that the cell interior is glassy but fluidized by the energy derived from metabolism. Our study then implies that there may be a positive feedback mechanism between mechanics and metabolism, where the fluidized cytoplasm in turn facilitates metabolic processes (11, 12, 19, 89, 90).

The active mechanics of cells were examined further in light of the non-thermal fluctuations. The force spectrum of the active non-thermal force was estimated from the FDT violation, based on the Langevin model extended to non-equilibrium systems. However, our analysis did not reveal a significant difference between normal and ATP-depleted cells, suggesting that the force spectrum is not indicative of the cell's metabolic activities. For one, it is important to note, that the efficiency of the energy conversion conducted by a mechano-enzyme depends on mechanical load. Presumably, the energy stored in ATP is more effectively converted to a mechanical form when the viscoelasticity of the surrounding cytoplasm is increased for ATP-depleted cells. For another, a non-equilibrium Langevin model that assumes microscopic activities is not relevant to describe MR in cells because active mechano-enzymes directly generate mesoscopic fluctuations.

We therefore proposed a mesoscopic continuum model with non-microscopic activities, instead of the non-equilibrium Langevin model.

In living cells with and without ATP depletion, the non-thermal fluctuations were found to be diffusion-like, or Markovian, and were designated as “active diffusion”. In contrast, the mechanical properties of cytoplasm were highly viscoelastic, making thermal fluctuations anomalous. This behavior cannot be explained if the mechano-enzyme activity and the fluctuating fields created in the surrounding medium were kept correlated. Rather, it is inferred that the mechano-enzyme activity enhances the structural relaxations (yielding) that generate large-scale fluctuations, as typically seen in a glassy medium. These stochastic, activity-induced relaxations erase the correlation between activity and fluctuations, resulting in active diffusion. Since the frequency of the local yielding is likely relates to the enzymatic activity, we conclude that the diffusivity and power of the non-thermal fluctuations reflects the enzymatic activity in the living cytoplasm.

## APPENDIX A: derivation of PSD in mesoscopic continuum model

In the mesoscopic continuum model, the non-thermal fluctuations of the probe particle are driven by a population of force dipoles distributed in continuum, as shown in Fig. 5E. The PSD of the probe particle was derived in a previous study for a two-component systems (57), the derivation and the result of which were complicated more than necessary. We then summarize the version of single component incompressible medium whose mechanical properties is denoted solely by $G(\omega)$.

Assuming that the linear superposition of fluctuation is possible, the probe fluctuation is given by the sum of fluctuations driven by each force dipole, as given by $\boldsymbol{u}_{\mathrm{A}}(t)=\sum_{|\boldsymbol{r}_i|>a}\boldsymbol{u}_{\mathrm{A}}(t;-\boldsymbol{r}_i,\hat{\boldsymbol{n}}_i)$. Assuming that force dipoles are distributed over space with a number density $c(\boldsymbol{r},\hat{\boldsymbol{n}})$ in a medium, the probe particle at the origin is then displaced by

$$\boldsymbol{u}_{\mathrm{A}}(t)=\sum_i\boldsymbol{u}_{\mathrm{A}}(t;-\boldsymbol{r}_i,\hat{\boldsymbol{n}}_i)\cong\int_{x>a}d\boldsymbol{r}\int d\hat{\boldsymbol{n}}\boldsymbol{u}_{\mathrm{A}}(t;-\boldsymbol{r},\hat{\boldsymbol{n}})c(\boldsymbol{r},\hat{\boldsymbol{n}}). \tag{A1}$$

By assuming that mechano-enzymes are randomly distributed and oriented, the numberdensity in space, $c(\boldsymbol{r})=\int d\hat{\boldsymbol{n}}c(\boldsymbol{r},\hat{\boldsymbol{n}})\equiv\sum_i\delta(\boldsymbol{r}-\boldsymbol{r}_i)$, satisfies the condition $\langle c(\boldsymbol{x})c(\boldsymbol{y})\rangle=c\delta(\boldsymbol{x}-\boldsymbol{y})$ where $c$ is the mean number density. Also by assuming that the mechano-enzymes work independently, the PSD of the probe particle’s active fluctuation is obtained by $\langle|\tilde{u}_{\mathrm{A}}(\omega)|^2\rangle=\frac{c}{3}\int_{r>a}d\boldsymbol{r}\int d\hat{\boldsymbol{n}}|\tilde{\boldsymbol{u}}_{\mathrm{A}}(\omega;-\boldsymbol{r},\hat{\boldsymbol{n}})|^2$. The actual calculation was conducted in Fourier space for brevity as shown in (57), such that

$$\langle|\tilde{u}_{\mathrm{A}}(\omega)|^2\rangle=c\langle|\tilde{\kappa}(\omega)|^2\rangle/90\pi a|G|^2. \tag{A2}$$

Here, we performed a Fourier transform of Eq. (14) with $t$, and used the Stokes relation [Eq. (1)] before integrating the contribution of independent force dipoles in the medium. With the same assumptions that lead us to Eq. (A2), we obtain,

$$\omega^2\langle|\tilde{u}_{\mathrm{A}}(\omega)|^2\rangle=\langle|\tilde{v}_{\mathrm{A}}(\omega)|^2\rangle\sim\frac{2cd^4}{9a}\left\langle\left|\tilde{\dot{\Delta}}(\omega)\right|^2\right\rangle \tag{A3}$$

## AUTHOR CONTRIBUTION

K.U., K.N., W.N., S.I. and H.E. collected/analyzed MR data. H.E., Y.S., and D.M. discussed/analyzed the results. D.M. designed/supervised the project, and wrote the manuscript with supports of H. E. and Y.S.

## ACKNOWLEDGEMENTS

This work was supported by JSPS KAKENHI Grant Numbers JP21H01048, JP20H05536, JP20H00128. We thank N. Oyama in Toyota Central R&D Labs for providing numerical simulation data, and A. Ikeda in Univ. Tokyo, K. Kanazawa in Kyoto University for helpful discussions.

**CONFLICTS OF INTEREST** The authors declare no conflict of interest.